\documentclass[apj]{emulateapj}

\usepackage{array}
\usepackage{ulem}
\usepackage{epsf, graphicx, subfigure}
\usepackage{amsmath} 
\usepackage{longtable}     
\usepackage{natbib}
\usepackage[usenames]{color}
\usepackage{float}
\usepackage{hyperref, url}

\definecolor{khaiorange}{RGB}{204,85,0}

\begin{document}

\title{Likelihood for Detection of Sub-parsec Supermassive Black Hole Binaries in Spectroscopic Surveys}
\author{Bryan J. Pflueger\altaffilmark{1}, Khai Nguyen\altaffilmark{1}, Tamara Bogdanovi\'c\altaffilmark{1}, Michael Eracleous\altaffilmark{2}, Jessie C. Runnoe\altaffilmark{3}, Steinn Sigurdsson\altaffilmark{2}, \& Todd Boroson\altaffilmark{4}}

\email{tamarab@gatech.edu, khainguyen@gatech.edu}
\altaffiltext{1}{Center for Relativistic Astrophysics, School of Physics, Georgia Institute of Technology}
\altaffiltext{2}{Department of Astronomy and Astrophysics, The Pennsylvania State University}
\altaffiltext{3}{Department of Astronomy, University of Michigan}
\altaffiltext{4}{Las Cumbres Observatory}

\begin{abstract}
Motivated by observational searches for sub-parsec supermassive black hole binaries (SBHBs) we develop a modular analytic model to determine the likelihood for detection of SBHBs by ongoing spectroscopic surveys. The model combines the parametrized rate of orbital evolution of SBHBs in circumbinary disks with the selection effects of spectroscopic surveys and returns a multivariate likelihood for SBHB detection. Based on this model we find that in order to evolve into the detection window of the spectroscopic searches from larger separations in less than a Hubble time, $10^8M_\odot$ SBHBs must, on average, experience angular momentum transport faster than that provided by a disk with accretion rate $0.06\,\dot{M}_E$. Spectroscopic searches with yearly cadence of observations are in principle sensitive to binaries with orbital separations $< {\rm few}\times 10^4\, r_g$ ($r_g = GM/c^2$ and $M$ is the binary mass), and for every one SBHB in this range there should be over 200 more gravitationally bound systems with similar properties, at larger separations. Furthermore, if spectra of all SBHBs in this separation range exhibit the AGN-like emission lines utilized by spectroscopic searches, the projection factors imply five undetected binaries for each observed $10^8M_\odot$ SBHB with mass ratio $0.3$ and orbital separation $10^4\,r_g$ (and more if some fraction of SBHBs is inactive).  This model can be used to infer the most likely orbital parameters for observed SBHB candidates and to provide constraints on the rate of orbital evolution of SBHBs, if observed candidates are shown to be genuine binaries.
\end{abstract}



\section{Introduction}
\label{sec:Intro}

Galactic mergers are an important driver of galaxy evolution and a natural channel for formation of supermassive black hole (SBH) pairs and possibly, multiplets \citep{Begelman1980}. Most of the scientific attention was initially focused on the galaxies in the act of merging, and their central SBHs were considered passive participants, taken for a ride by their hosts. The  realization that SBHs play an important role in the evolution of their host galaxies \citep{ferrarese00,gebhardt00,tremaine02} spurred a number of observational and theoretical studies which aim to determine what happens to SBHs once their host galaxies merge.

From an observational point of view, this question is pursued through electromagnetic searches for dual and multiple SBHs with a variety of separations, ranging from $\sim$10s kpc to sub-parsec scales. The multi-wavelength searches for SBH systems with large separations, corresponding to early stages of galactic mergers, have so far successfully identified a few dozen dual and offset active galactic nuclei \citep[AGN;][and others]{komossa03, koss11, koss16, liu13b, comerford15, barrows16}.

SBHs with even smaller (parsec and sub-parsec) separations are representative of the later stages of galactic mergers in which the two SBHs are sufficiently close to form a gravitationally bound pair. Throughout this paper, we refer to these as {\it supermassive black hole binaries} (SBHBs). Observational techniques used to search for such systems have so far largely relied on direct imaging, photometry, and spectroscopic measurements \citep[see][for a review]{Bogdanovic2015}. They have recently been complemented by observations with pulsar timing arrays (PTAs). We summarize the outcomes of these different observational approaches in the next few paragraphs and note that in all cases SBHBs have been challenging to identify because of their small separation on the sky, as well as the uncertainties related to the uniqueness of their observational signatures.  

Angular separation of the parsec and sub-parsec binaries is below the spatial resolution of most astronomical observations, except for radio observations using the Very Long Baseline Interferometry (VLBI) technique. This approach has been used to identify the most convincing SBHB candidate thus far, 0402+379 \citep{rodriguez06,rodriguez09,morganti09}, which hosts a pair of two compact radio cores at a projected separation of 7.3~pc on the sky. Recently, \citet{bansal17} reported
that long term VLBI observations reveal relative motion of the two cores consistent with orbital motion, lending further support to the SBHB hypothesis. Apart from this serendipitously discovered system, direct imaging with VLBI did not reveal many others \citep{burke11,condon11}, largely because this technique cannot be used in survey mode, and so it requires a prior knowledge of likely candidates. 

Photometric surveys, like the Catalina Real-Time Transient Survey, the Palomar Transient Factory and others, have  uncovered about 150 SBHB candidates \citep[e.g.,][]{valtonen08, Graham2015, charisi16, liu16}. In this approach, the quasi-periodic variability in the lightcurves of monitored quasars is interpreted as a manifestation of binary orbital motion. Because of the finite temporal extent of the surveys, which must record at least several orbital cycles of a candidate binary, most photometrically identified SBHB candidates have relatively short orbital periods, of order a few years. However, irregularly sampled, stochastically variable lightcurves of ``normal" quasars, powered by a single SBH, can be mistaken for periodic sources and can lead to false binary identifications in photometric surveys \citep{Vaughan2016}. This possibility is supported by PTAs, which are capable of probing the gravitational wave background at nanoHertz frequencies from sources like SBHBs with orbital periods of a few years \citep{lentati15, shannon15, arzoumanian16}. Specifically, the upper limit placed by PTAs largely rules out the amplitude of gravitational wave background resulting from the $\sim150$ photometric binary candidates, implying that some significant fraction of them are unlikely to be SBHBs \citep{sesana17}. While this will lead to a downward revision in the number of photometrically identified SBHB candidates, it provides a nice example of the effectiveness of multi-messenger techniques, when they can be combined.

Spectroscopic searches for SBHBs have so far identified several dozen candidates. They rely on the detection of a velocity shift in the emission-line spectrum of an SBHB that arises as a consequence of the binary orbital motion. This approach is similar to that for detection of single- and double-line spectroscopic binary stars, where the lines are expected to oscillate about their local rest-frame wavelength on the orbital time scale of the system. In the context of the binary model, the spectral emission lines are assumed to be associated with gaseous accretion disks that are gravitationally bound to the individual SBHs \citep{gaskell83,gaskell96,bogdanovic09}. The main complication of this approach is that the velocity-shift signature is not unique to SBHBs \citep[e.g.,][]{popovic12,Eracleous2012,barth15}. To address this ambiguity, spectroscopic searches have been designed to monitor the offset of broad emission-line profiles over multiple epochs and target sources in which modulations in the offset are consistent with binary orbital motion \citep{bon12, bon16, Eracleous2012, decarli13, Ju2013, liu13, shen13, Runnoe2015, Runnoe2017, li16, wang17}. Two drawbacks of this approach however remain: (a) temporal baselines of spectroscopic campaigns cover only a fraction of the SBHB orbital cycle and (b) intrinsic variability of the line profiles may mimic or hide radial velocity variations \citep[see Appendix A of][]{Runnoe2017}.

In this work, we present an analytic model that can be used to calculate what subset of SBHBs is detectable by spectroscopic searches, given their underlying assumptions and selection effects. We also examine what can be learned about this population if a sample of bona fide SBHBs is detected. The rest of the paper is organized as follows. In \S~\ref{sec:ModelMethod} we review the selection effects of spectroscopic searches and assumptions adopted by our model. In \S~\ref{sec:results} we present a calculation of the probability density functions and in \S~\ref{sec:results2} the likelihood for detection of a sub-parsec SBHB. We discuss the implications of our findings in \S~\ref{sec:discussion} and conclude in \S~\ref{sec:conclusions}.


\section{Model assumptions and parameters}
\label{sec:ModelMethod}


\subsection{Selection effects in spectroscopic searches for SBHBs}
\label{sec:selection}

In this section we review the assumptions commonly made by the optical spectroscopic searches for sub-parsec SBHBs and summarize their selection effects. Throughout this paper we consider the observational campaign presented in \citet{Eracleous2012} and \citet{Runnoe2015,Runnoe2017} as an illustrative example and note that similar considerations can be applied to other spectroscopic studies. Hereafter, we refer to it as the E12 spectroscopic campaign.

The principal assumption made by all spectroscopic searches is that some fraction of SBHBs at sub-parsec orbital separations are contained in a region comparable to or larger in size than the broad line regions (BLRs) of regular AGNs that do not host SBHBs. If the emission properties of BLRs that host SBHBs remain qualitatively similar to the single AGN case, it follows that their dynamical perturbation by SBHBs can in principle be reflected in the low-ionization broad emission line profiles \citep[e.g.,][]{gaskell83,gaskell96,bogdanovic08, montuori11, montuori12}. The broad emission lines of particular interest are H$\alpha~\lambda6563$, H$\beta\;\lambda4861$ and Mg\,\textsc{ii}$\;\lambda2798$, because they are prominent in AGN spectra and are commonly used as tracers of dense, low-ionization gas in BLRs at low (H$\alpha$ at z $<$ 0.4) and high redshift (Mg\,\textsc{ii} at z $<$ 2.5).

An additional assumption adopted by spectroscopic searches for SBHBs is that the flux in the broad emission line is dominated by the portion of the gas flow bound to the secondary, lower mass SBH. This assumption is directly motivated by a number of theoretical studies of SBHBs in circumbinary disks \citep[e.g.,][]{Armitage2005, macfadyen08}. These studies show that binary torques can truncate a sufficiently cold circumbinary accretion flow and create an inner, low density cavity by evacuating the gas from the central portion of the circumbinary disk \citep{lp79}. SBHs in this phase accrete by capturing the gas from the inner rim of the circumbinary disk and in this way can maintain mini-disks bound to individual holes. Hydrodynamic simulations of prograde binaries (rotating in the same sense as the circumbinary disk) indicate that in unequal mass binaries accretion occurs preferentially onto the smaller of the two objects, which orbits closer to the inner edge of the circumbinary disk \citep{Artymowicz1994, gunther02, hayasaki07, roedig11, farris14}. Taken at face value, this suggests that the AGN associated with the secondary SBH may be more luminous than the primary, making the system analogous to single-line spectroscopic binary stars. Following the approach of the spectroscopic surveys, we too adopt this assumption in the model presented here. This assumption does not change the observational results of the spectroscopic surveyes, only their inference about which of the SBHs is active. We also note that recent theoretical models indicate that the contribution to the optical broad emission lines by the BLR bound to the primary SBH is not negligible and can be dominant in the majority of SBHB configurations \citep{Nguyen2016}.

With these assumptions, the E12 campaign searched for $z < 0.7$ SDSS\footnote{Sloan Digital Sky Survey} AGN with broad H$\beta$ lines offset from the rest frame of the host galaxy by $\gtrsim 1000\,{\rm km\,s^{-1}}$. In the context of the SBHB model described above, this velocity offset corresponds to the velocity of the secondary SBH projected along the observer's line of sight. Based on this criterion, E12 selected 88 quasars for observational followup from an initial group of about 16,000 objects. The followup observations span a temporal base line from few weeks to 12 years in the observer's frame. Their goal is to measure the epoch-to-epoch modulation in the velocity offset of the H$\beta$ profiles and to test the binarity hypothesis. The relative velocity of the broad H$\beta$ profiles between different epochs has been measured with an uncertainty of $\lesssim 40\,{\rm km\,s^{-1}}$ for 80\% of the sample and $\lesssim 55\,{\rm km\,s^{-1}}$ for the entire sample. After multiple epochs of followup, statistically significant changes in the velocity offset have been measured in 29/88 candidates and reported in the publications mentioned above. 

We therefore adopt two criteria, $V_{\rm lim} = 1000\,{\rm km\,s^{-1}}$ for the initial velocity offset and $\Delta V_{\rm lim} = 40\,{\rm km\,s^{-1}}$ for velocity modulations, to describe the selection effects of the E12 survey. It is worth noting that some spectroscopic campaigns do not impose a cut in the initial velocity offset and consider as SBHB candidates all AGN for which $\Delta V_{\rm lim}$ is not zero and is consistent with the binary orbital motion \citep{shen13, Ju2013, liu13, wang17}. The advantage of the latter approach is that it starts with a larger statistical sample of AGN which are searched for apparent radial velocity variaitons. But the more stringent selection criteria in the former approach reduce the chance of confusion with regular AGN, the majority of which are characterized by $V_{\rm lim} < 1000\,{\rm km\,s^{-1}}$. In this case however it may take longer to detect radial velocity variations, since for circular orbits the SBHBs spend most of their time at the highest velocity offsets, where the projected acceleration is the lowest.


\subsection{Key parameters of the model}
\label{S_params}

We describe each SBHB configuration in terms of four intrinsic parameters: the total mass of the binary ($M= M_1+M_2$), its mass ratio ($q = M_2/M_1 \leq 1$), the orbital separation ($a$), and the effective accretion rate through the disk ($\dot{M} = \dot{M_1} + \dot{M_2}$).  We only consider SBHBs on circular orbits for simplicity and provide constraints and considerations relevant for each parameter below.

We consider binaries with a mass between $10^7\,M_\odot$ and $10^9\,M_\odot$ and a range of values for $q$ that is motivated by simulations of galaxy mergers that follow pairing of their SBHs. These show that SBH pairs with mass ratios $q<0.1$ are less likely to form gravitationally bound binaries within a Hubble time at any redshift, primarily due to the inefficiency of dynamical friction on the lower mass SBH \citep{callegari09,callegari11}. They also find that SBH pairs with initially unequal masses tend to evolve towards equal masses, through preferential accretion onto a smaller SBH. This trend is also consistent with that found by \citet{Kelley2017}, based on the more recent Illustris simulation. We therefore adopt $0.1 \leq q \leq 1$ in this model, leading to a range of masses for the secondary SBHs, of $10^6\,M_\odot \lesssim M_2 \leq 10^9\,M_\odot$. This range is consistent with the masses of SBHs powering the regular AGN, not hosting SBHBs, commonly observed by SDSS and similar ground based, spectroscopic surveys. 

We further consider a range of orbital separations that characterize gravitationally bound SBHBs. Qualitatively, a gravitationally bound SBHB forms when the amount of gas and stars enclosed within its orbit becomes comparable to $M$.  This orbital separation is comparable to the radius of gravitational influence of a single SBH, where the circular velocity around the black hole equals the stellar velocity dispersion, $r_{\rm inf} = GM/ \sigma_*^2$.  Combining this with the $M-\sigma_*$ relationship reported in \citet{mcconnell13} we obtain
\begin{equation}
r_{\rm inf} \approx 14 \, M_8^{0.645}\,{\rm pc} = 2.9\times10^{6}\, r_g \, M_8^{-0.355}\; .
\end{equation}
where $r_g = GM/c^2$ is the gravitational radius and $M_8 = M/10^8M_\odot$. We therefore assume that gravitationally bound binaries form when $a_{\rm max} \sim 10^6\,r_g$ for a wide range of SBHB masses. In this work we follow their evolution from $a_{\rm max}$ to $a_{\rm min} = 10^2\,r_g$, below which the tidal truncation is likely to render the BLR of the secondary SBH too compact to emit prominent broad optical emission lines.

For each SBHB configuration we parametrize the rate of orbital evolution of the binary using the effective accretion rate, $0.01\dot{M}_E \leq \dot{M} \leq \dot{M}_E$.  Here $\dot{M}_E = L_E/\eta c^2$ is the Eddington accretion rate, $\eta$ is the radiative efficiency, $L_E = 4\pi GM m_p c/\sigma_T$ is the Eddington luminosity, $\sigma_T$ is the Thomson cross section and other constants have their usual meaning. For an SBHB embedded in the circumbinary disk, $\dot{M}$ corresponds to the accretion rate through the geometrically thin and optically thick gas disk described by the Shakura-Sunyaev $\alpha$-disk model \citep{Shakura1973}. 

\begin{figure*}[t]
\centerline{
\includegraphics[trim=0 0 0 0, clip, scale=0.75,angle=0]{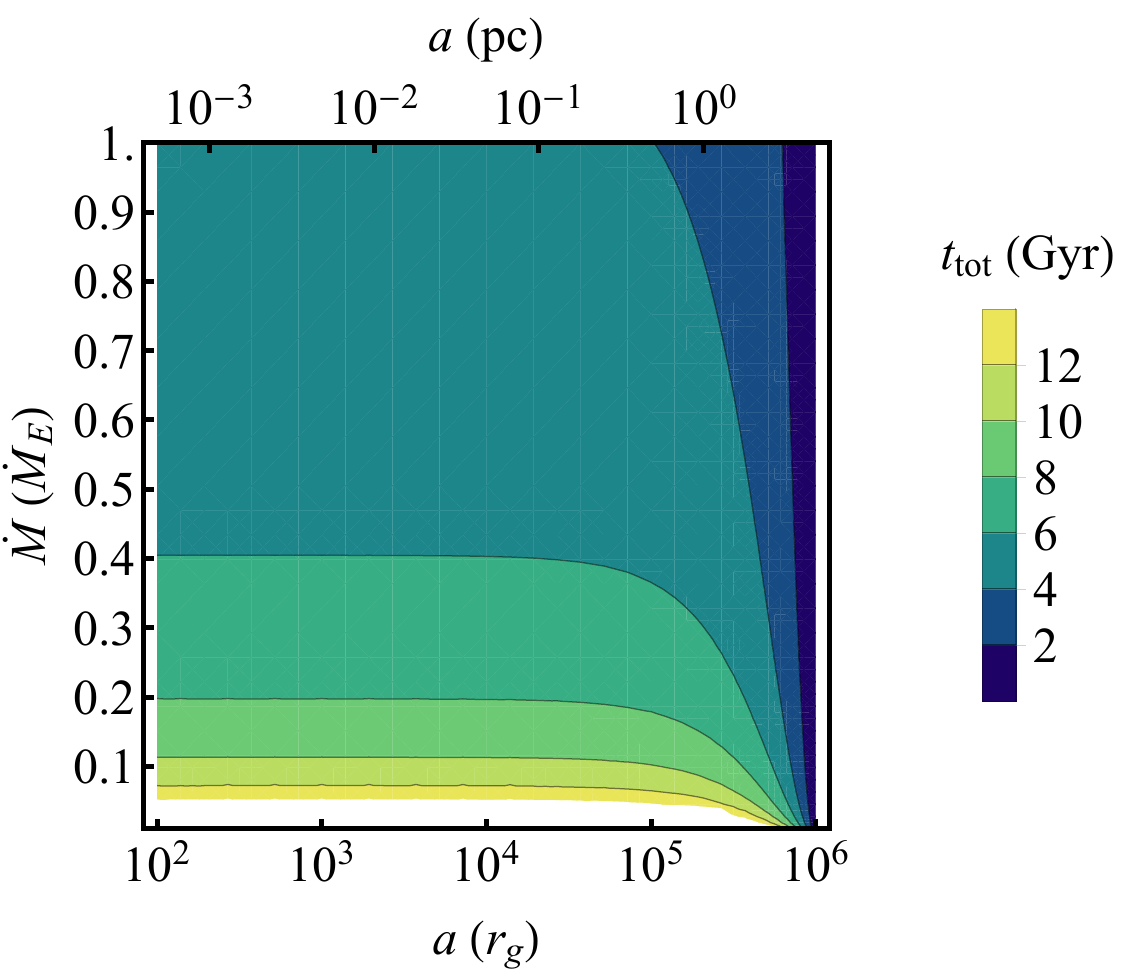} 
\includegraphics[trim=-20 0 0 0, clip, scale=0.75,angle=0]{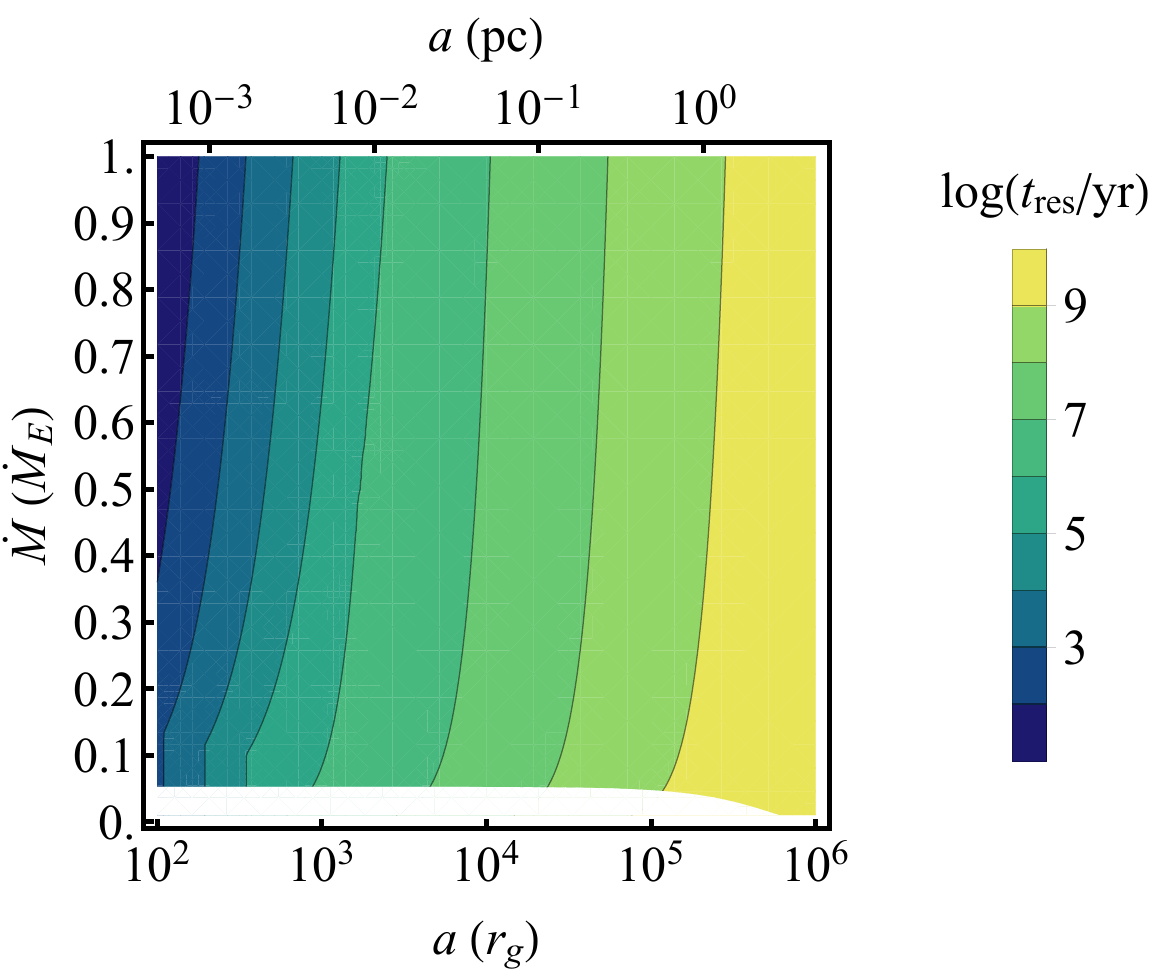} 
}
\caption{{\it Left:} Cumulative time ($t_{\rm tot}$) for an SBHB with $M=10^8M_\odot$ and $q=0.3$ to evolve from the point of formation of a gravitationally bound pair to separation {\it a}. The white region at the bottom of the plot marks SBHB configurations for which $t_{\rm tot}$ exceeds a Hubble time. {\it Right:} Residence time at a given separation, $a/\dot{a}$, for an SBHB with the same $M$ and $q$ as in the left panel.} 
\label{timeplots}
\end{figure*}

It is worth noting that the Shakura-Sunyaev solution for an accretion disk around a single SBH implies that for each $M$ and $\dot{M}$ there is a critical radius beyond which the disk is gravitationally unstable to fragmentation and star formation. This truncation radius in general arises somewhere between $10^2-10^6\,r_g$, depending on the properties of the disk and the mass of the SBH \citep[see equations~15 and 16 in][for example]{haiman09}. This is commonly considered as the outer edge of the accretion disk beyond which the transport of angular momentum transitions from accretion torques to some other mechanism, either stellar or gaseous. Given that the structure and stability of circumbinary accretion disks is an area of active research, the existence and location of such outer truncation radius are still open questions. In light of this fundamental uncertainty we adopt $\dot{M}$ as a proxy for the rate of angular momentum transport by any physical mechanism and do not require the presence of a stable circumbinary disk on all scales. Specifically, we use it to describe the rate of angular momentum transport that would correspond to that of a disk with an accretion rate $\dot{M}$ around a single SBH with the mass $M$, equal to that of the SBHB. 

In summary, we use the following parameters to describe the properties of the SBHB and selection effects of the E12 spectroscopic survey:
\begin{eqnarray}\nonumber
10^7M_\odot \leq && M  \leq 10^9M_\odot \\ \nonumber
0.1\leq && q \leq1 \\  \nonumber
10^2 r_g \leq  &&  a  \leq 10^6 r_g \\ \nonumber
0.01\,\dot{M}_E\leq &&  \dot{M} \leq \dot{M}_E \\ \nonumber
V_{\rm lim} = 1,000\,\rm{km \, s^{-1}} && {\rm and}\;\; \; \Delta V_{\rm lim} = 40 \ \rm{km \, s^{-1}}  \nonumber
\end{eqnarray}
%

\section{Intrinsic Probability Density Functions}
\label{sec:results}
\subsection{Probability of a binary residing at separation $a$: $\rho(a)$}
\label{subsec:orbitevol}

For each configuration we evaluate the rate of orbital evolution of the binary and a probability that it resides at some orbital separation, $a$. The total probability of finding the SBHB anywhere between the formation radius and coalescence can be defined as
\begin{equation}
\int_{a_{\rm max}}^{0} \rho(a) \ da = \int_{10^6 r_g}^{0} \frac{da}{\dot{a}}\frac{1}{t_{\rm tot}} \equiv 1\; ,
\label{eq_rhoa}
\end{equation}
where $\rho(a)$ is the probability density and $t_{tot}$ is the total time it takes the SBHB to evolve from $a_{\rm max}$ to coalescence. Because the rate of orbital evolution, $\dot{a}$, is a function of $M$, $q$ and $\dot{M}$ it follows that $\rho(a) \equiv \rho(a|M,\dot{M},q)$. Hereafter, we adopt a simplifying assumption that the values of $M$, $\dot{M}$, and $q$ are constant throughout the evolution of an SBHB, and discuss the implications in \S~\ref{S_simplifications}. 

If the orbital evolution of the binary is driven by a circumbinary disk, the rate of shrinking of the SBHB orbit can be described in terms of the viscous inflow rate at the disk inner edge as 
\begin{equation}
\dot{a}_{\rm visc}=-\frac{3}{2}\alpha \left( \frac{h}{r}  \right)^2V_{\rm Kep}
\end{equation}
where $\alpha = 0.1$ is the viscosity parameter, $h/r$ is the aspect ratio of the disk, and $V_{\rm Kep} = (2a/r_g)^{-1/2}$ is the circular orbital speed of the circumbinary disk with an inner edge at the radius $r = 2a$ \citep{macfadyen08}. Using the expression for $h/r$ by \citet{Shakura1973}, we obtain the infall rate for the gas pressure (outer) and radiation pressure dominated (inner)  portion of the disk:
\begin{eqnarray}
\dot{a}_{\rm gas} &=& -4.24\times10^{-6}c\,\, \alpha^{4/5}\,\dot{m}^{2/5}\,\tilde{a}^{-2/5} M_8^{-1/5} \label{eq_adot_g}\; ,\\
\dot{a}_{\rm rad} &=& -1.21\times10^{2}c\,\, \alpha\, \dot{m}^{2}\,\tilde{a}^{-5/2}\; .
\label{eq_adot_r}
\end{eqnarray}
where we introduce dimensionless parameters $\dot{m} = \dot{M}/\dot{M}_E$ and $\tilde{a} = a/r_g$, so that $\dot{a}$ is expressed in terms of the speed of light. SBHBs whose orbital evolution is driven by the emission of gravitational waves (GWs) shrink at the rate given by the expression from \citet{Peters1964} for circular orbits
\begin{equation}
\dot{a}_{\rm gw}=-12.8c\,\,\frac{q}{(1+q)^2}\tilde{a}^{-3} \;.
\label{eq_adot_gw}
\end{equation}
We use the above expressions for the rate of orbital shrinking to calculate the time that a gravitationally bound SBHB spends evolving through each regime 
\begin{equation}
t_{\rm x}=\int_{a_{\rm x,i}}^{a_{\rm x,f}}\frac{da}{\dot{a}_{\rm x}}\; ,
\label{eq_tx}
\end{equation}
where ``x'' stands for ``gas", ``rad" or ``gw", and $a_i$ and $a_f$ are the initial and final orbital separations which determine the boundaries of a particular regime, as defined in Appendix~\ref{app:radii}. After evaluating the integral in equation~\ref{eq_tx} for constant $M$, $q$ and $\dot{M}$ we obtain
\begin{eqnarray}
 t_{\rm gas} &=& 8.3\times10^7\,{\rm s}\,\, \alpha^{-4/5} \dot{m}^{-2/5}
M_8^{6/5} \left[\tilde{a}_{\rm gas,i}^{7/5}-\tilde{a}_{\rm gas,f}^{7/5} \right],\\
t_{\rm rad} &=& 1.2\,{\rm s}\,\,\alpha^{-1}\,\dot{m}^{-2} M_8 \left[\tilde{a}_{\rm rad,i}^{7/2}-\tilde{a}_{\rm rad,f}^{7/2} \right],\; {\rm and}\\
t_{\rm gw} &=& 9.6\,{\rm s}\,\, \frac{(1+q)^2}{q}\, M_8 \left[\tilde{a}_{\rm gw,i}^4-\tilde{a}_{\rm gw,f}^4  \right]\, .
\end{eqnarray}
The total time to coalescence is given by
\begin{equation}
t_{\rm tot}=t_{\rm gas}+t_{\rm rad}+t_{\rm gw} \;.
\label{eq_ttot}
\end{equation}

The left panel of Figure~\ref{timeplots} illustrates the total (cumulative) time for orbital evolution of an SBHB with mass $10^8M_\odot$ and $q=0.3$ from the instance it becomes gravitationally bound to some orbital separation $a < a_{\rm max}$. The time scale is longest for SBHB configurations in which the transport of orbital angular momentum is inefficient, as indicated by low $\dot{M}$. Note that for $\dot{M} < 0.1\,\dot{M}_E$ the total time for evolution exceeds a Hubble time, as marked by the white region at the bottom of the plot. It follows that $10^8M_\odot$ SBHBs evolve into the observational window of spectroscopic surveys within the age of the universe if the transport of angular momentum (due to all mechanisms) is equivalent to a  disk with $\dot{M} \gtrsim 0.1\,\dot{M}_E$.

%
\begin{figure}[t]
\begin{center}
\includegraphics[scale=.41]{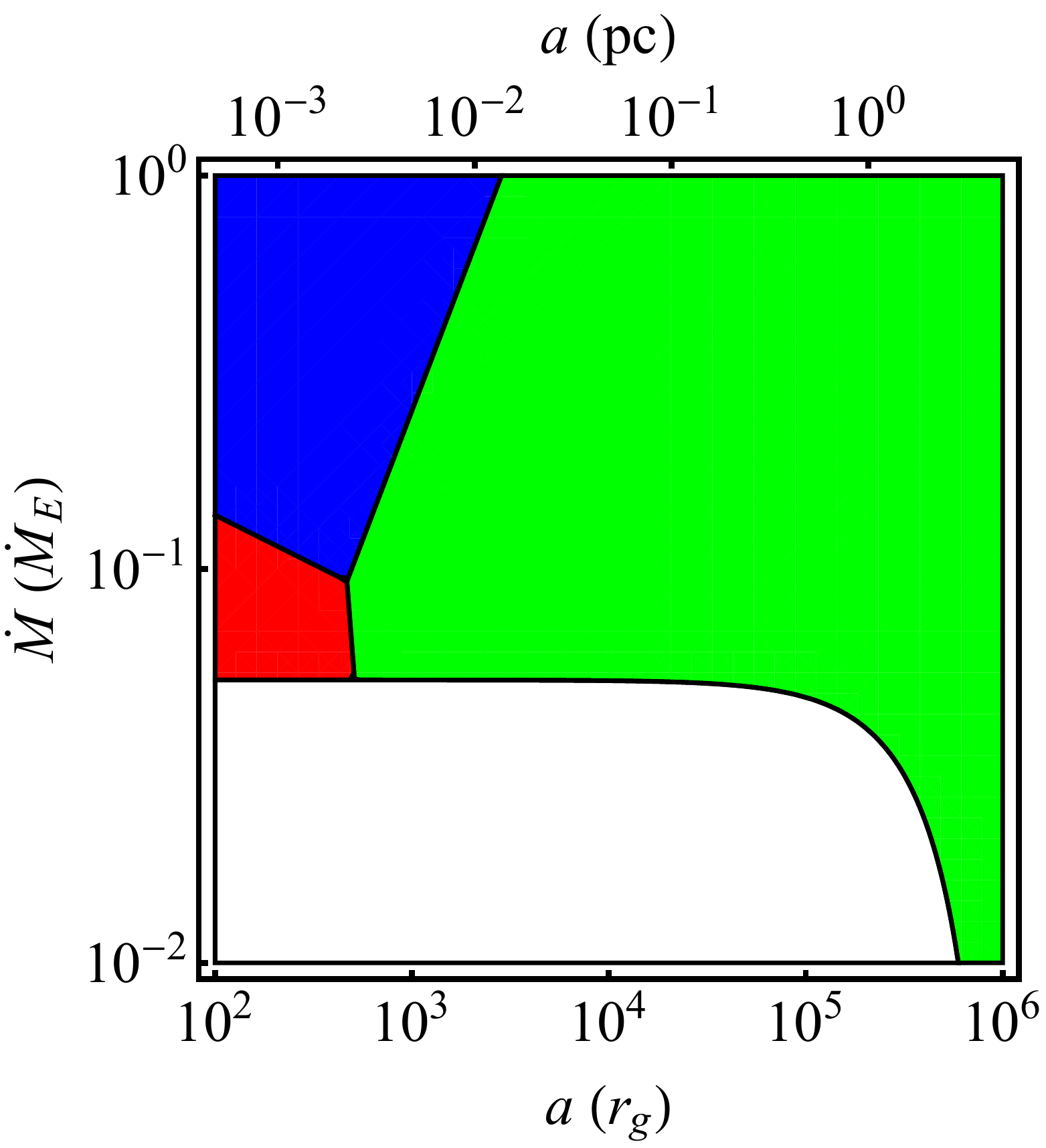} 
\caption{Regimes of binary orbital evolution calculated for $M=10^8M_\odot$ and $q=0.3$. Different colors mark regions of the parameter space in which the SBHB orbit shrinks due to torques in the gas pressure dominated (green), radiation pressure dominated portion of the circumbinary disk (blue), and due to GW emission (red). The white region corresponds to SBHB configurations for which evolution times exceed a Hubble time.}
\label{fig:regimes}
\end{center}
\end{figure}
%
\begin{figure}[t]
\begin{center}
\includegraphics[trim=0 -10 0 0, clip, scale=0.76,angle=0]{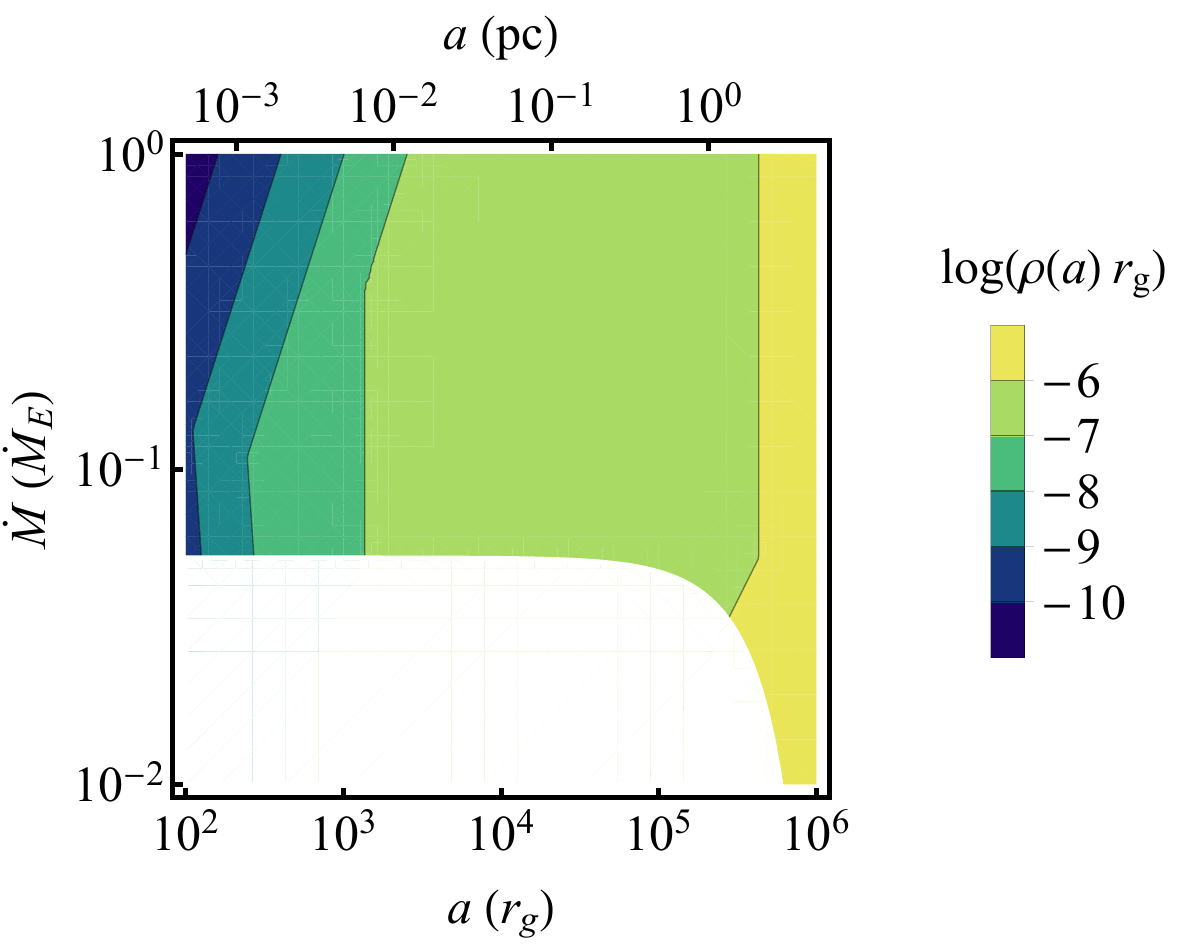}  
\includegraphics[trim=-10 0 0 0, clip, scale=0.75,angle=0]{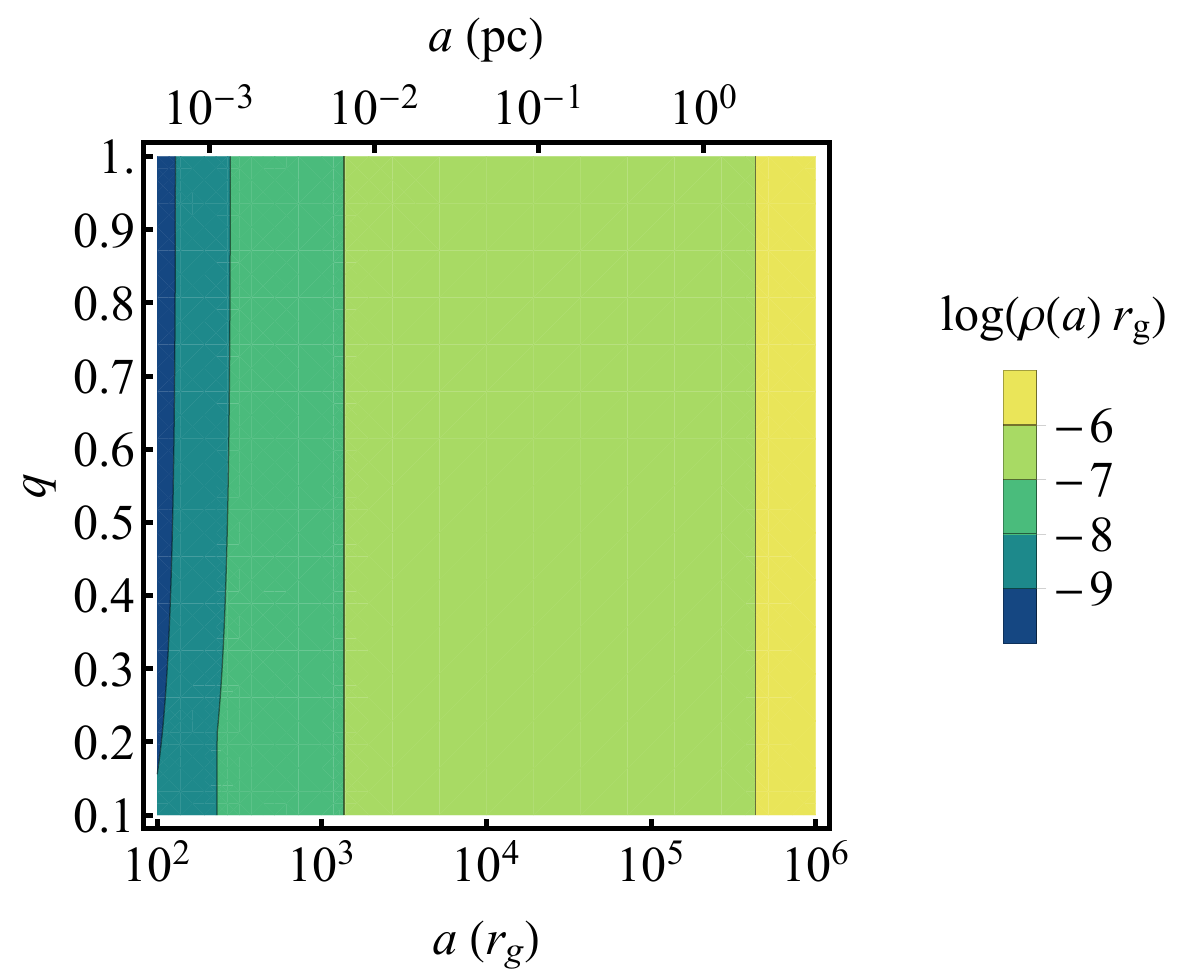}   
\end{center}
\caption{Two-dimensional cuts of the multi-variate probability density function, $\rho(a|M,\dot{M},q)$, plotted for $M=10^8M_\odot$ and $q=0.3$ (top) and $\dot{M}=0.1\dot{M}_E$ (bottom). In this model, the majority of gravitationally bound SBHBs reside at larger orbital separations and their orbital evolution exhibits only a weak dependance on $q$. Color marks the probability density on a logarithmic scale, $\log (\rho(a)\,r_g)$. The white region in the top panel corresponds to SBHB configurations for which evolution times exceed a Hubble time.}
\label{rho_a_app}
\end{figure}

The right panel of Figure \ref{timeplots} illustrates the characteristic residence time of the SBHB at a given separation, calculated as $a/\dot{a}$, and evaluated for the same $M$ and $q$ as the left panel. It shows that binaries tend to spend more time at large orbital separations and evolve more rapidly at small separations, assuming an uninterrupted transport of orbital angular momentum.

It is also interesting to examine which mechanism dominates the transport of orbital angular momentum, by color coding different regimes of $\dot{a}$ as a function of $\dot{M}$ and $a$. Figure~\ref{fig:regimes} shows the portions of the parameter space in which the SBHB orbit shrinks due to torques in the gas pressure dominated (green), radiation pressure dominated portion of the circumbinary disk (blue) or due to GW emission (red). 

Figure~\ref{rho_a_app} shows two-dimensional cuts of the multi-variate probability density for an SBHB to reside at an orbital separation $a$, as defined in equation~\ref{eq_rhoa}. It is worth noting that when $\dot{a}$ is expressed in terms of the speed of light and $t_{\rm tot}$ is evaluated in units of $r_g/c$ (instead of seconds), $\rho(a)$ acquires units of $r_g^{-1}$ and does not explicitly depend on the mass of the SBHB\footnote{The implicit dependence on the SBHB mass however remains because $r_g$ is a function of $M$.}. In other words, probability density expressed in these units is of the same magnitude, regardless of the SBHB mass. We adopt these geometric units in plots of the probability density in Figure~\ref{rho_a_app}.

The top panel of Figure~\ref{rho_a_app} shows that the majority of gravitationally bound SBHBs should reside at larger orbital separations, where they spend most of their life according to Figure~\ref{timeplots}. The contour breaks in this panel mimic those in Figure~\ref{fig:regimes}, and effectively demarcate transitions between various regimes of orbital evolution. The bottom panel of  Figure~\ref{rho_a_app} illustrates the dependance of $\rho(a)$ on the mass ratio $q$. It shows that the evolution of the SBHB through the circumbinary disk does not depend on $q$, while the evolution in the GW regime only exhibits a weak dependance on $q$. This follows directly from our description of $\dot{a}_{\rm gas}$, $\dot{a}_{\rm rad}$ and $\dot{a}_{\rm gw}$ in equations~\ref{eq_adot_g}--\ref{eq_adot_gw}. 

Note that the rate of evolution through the circumbinary disk can, in principle, be a function of $q$, when the mass of the circumbinary disk is smaller than the mass of the secondary SBH and the circumbinary disk is the sole driver of orbital evolution (a.k.a., the secondary-dominated regime), as discussed in \citet{haiman09} and \citet{Rafikov2013} and adopted in the study by \citet{Ju2013}. Here we make a different assumption, that even if the size of the circumbinary disk is finite, some other mechanism takes over angular momentum transport beyond the outer truncation radius, resulting in the steady orbital evolution of the SBHB described by some characteristic $\dot{M}$, regardless of the binary mass ratio. We discuss the implications of this assumption in Section~\ref{S_simplifications}.


\subsection{Mass ratio probability distribution: $\rho(q)$}
\label{qprob}

\begin{figure}[t]
\begin{center}
\includegraphics[scale=.6]{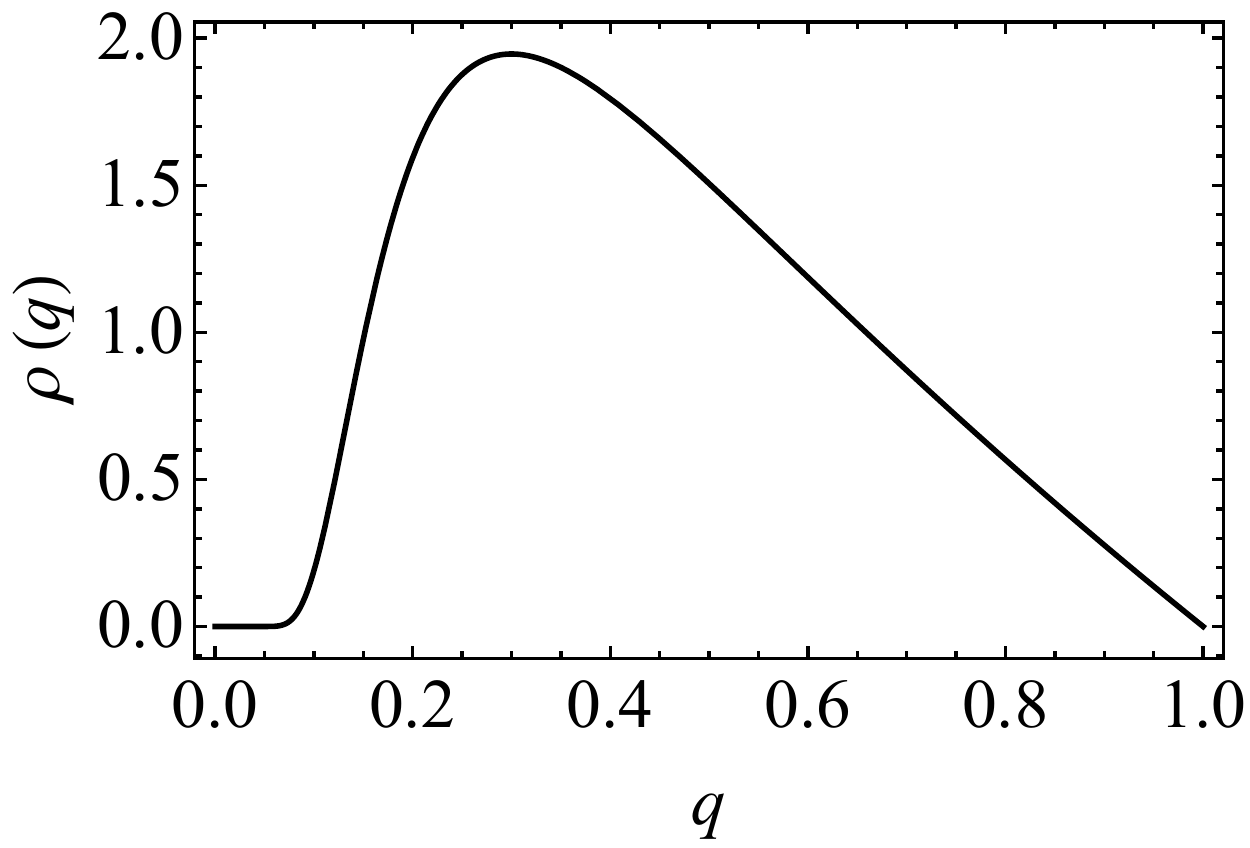} 
\end{center}
\caption{Probability distribution function for SBHB mass ratios adopted in this work. This is motivated by the results of cosmological simulations, as discussed in \S~\ref{qprob}.}
\label{massratiodis}
\end{figure}

Even though the orbital evolution of bound SBHBs in this model only weakly depends on $q$, it is still expected that the mass ratio of the SBH pairs that successfully form bound binaries are characterized by some initial distribution. This distribution results from the cosmological evolution of SBHs through galactic mergers and accretion in stages preceding the formation of a gravitationally bound binary. In order to account for this property we adopt an analytic expression for probability density loosely motivated by cosmological simulations that follow mergers of galaxies \citep{callegari09,callegari11, Kelley2017} and dark matter halos \citep{Stewart2009, Hopkins2010}
\begin{equation}
\rho(q)=\gamma\,q^{-0.3}\,(1-q)\,e^{-\beta/q^2}    
\label{eq_rhoq}
\end{equation}
where $\gamma=2.79$ and $\beta=3.28\times10^{-2}$ are the dimensionless parameters obtained from normalization of this distribution. For more details see Appendix~\ref{app:q}. 

Figure~\ref{massratiodis} illustrates the shape of the resulting distribution function, which peaks at $q\approx 1/3$. In reality the mass ratio distribution of SBHBs is a function of the total binary mass, redshift, and other parameters. We neglect these dependences for simplicity and note that because of the modularity of the model the above analytic expression can be readily replaced with a different prescription for $\rho(q)$. We discuss implications of these assumptions in \S~\ref{S_simplifications}.


\section{Results}
\label{sec:results2}

\subsection{Probability of observing a binary with a significant velocity offset: $P_V$}
\label{probvel}

The previous two sections describe the {\it intrinsic} probability that an SBHB exists at a certain orbital separation given the properties of the binary and some characteristic rate at which it evolves. In this section we discuss the ability of the spectroscopic searches to detect such binaries when they impose a selection criterion that the broad emission line profiles must be offset from the rest frame of the galaxy by some specified $V_{\rm lim} \neq 0$.

According to the assumptions adopted by E12 and similar surveys this velocity offset can be attributed to the orbital motion of the secondary SBH, with orbital speed relative to the center of mass of the binary, $V_2 = c/(1+q)\,\tilde{a}^{1/2}$. Projected along the observer's line of sight this orbital speed corresponds to an observed velocity offset of
\begin{equation}
V_{\rm obs}=\frac{1}{1+q}\frac{c}{\tilde{a}^{1/2}}\sin\phi\cos\theta
\label{lineofsightvelocity}
\end{equation}
where we define the angles in the coordinate system centered on the SBHB center of mass, so that the observer is located along the $x$-axis at infinity, $\phi$ defines the orbital phase of the secondary measured from the positive $x$-axis toward the positive $y$-axis, and $\theta$ is the angle between the orbital plane and the observer's line of sight.

\begin{table}[t]
\caption{Characteristic orbital separations} 
\centering 
\begin{tabular}{c c c  c c} 
\hline\hline 
$q$ & $M$ & $a_V^{\rm max}$ & $a_{\Delta V}^{\rm min}$ & $a_{\Delta V}^{\rm max}$    \\ [0.5ex] 
 & $(M_\odot)$ & $(10^3\,r_g)$ & $(10^2\,r_g)$ & $(10^3\,r_g)$ \\ [0.5ex] 
\hline 
0.1 & $10^7$ & 74 & 22 & 66  \\
0.1 & $10^8$ & 74 & 4.7 & 21  \\
0.1 & $10^9$ & 74 & 1.0 & 6.6 \\ 
0.3 & $10^7$ & 53 & 22 & 61  \\
0.3 & $10^8$ & 53 & 4.7 & 19  \\
0.3 & $10^9$ & 53 & 1.0 & 6.1 \\ 
1 & $10^7$ & 23 & 22 & 49  \\
1 & $10^8$ & 23 & 4.7 & 16  \\
1 & $10^9$ & 23 & 1.0 & 4.9 \\ 
\hline 
\end{tabular}
\tablecomments{$a_V^{\rm max}$ and $a_{\Delta V}^{\rm max}$ -- largest detectable values of SBHB orbital separation set by the selection effects of the E12 survey. $a_{\Delta V}^{\rm min}$ -- minimum orbital separation below which the model assumption $\Omega \Delta t \ll 1$ breaks down.}
\label{table:cutoff} 
\end{table}

The SBHB associated with this line-of-sight velocity can have a maximum orbital separation of
\begin{equation}
a_V^{\rm max}=\frac{9\times10^4\,r_g}{(1+q)^2} \left(\frac{V_{\rm obs}}{10^3\,{\rm km\,s^{-1}}} \right)^{-2} ,
\end{equation}
where we set $V_{\rm obs}$ equal to $V_{\rm lim} = 1000\, {\rm km\,s^{-1}}$. Table~\ref{table:cutoff} illustrates the values of $a_{V}^{\rm max}$ set by this cutoff for several different SBHB configurations. It is worth noting that, if the spectroscopic surveys instead of the secondary provide a measurement of the radial velocity of the primary SBH, the value of $a_{V}^{\rm max}$ would be smaller by a factor of $q^2$.

We infer the probability for detection of an SBHB by placing the observer on a sphere, centered on the SBHB, and by considering all SBHB configurations where the angles $\theta$ and $\phi$ are such that $V_{\rm obs} \geq V_{\rm lim}$ and the velocity offset from the binary motion is detected. The probability of detection is then given by the ratio of the surface area where this condition is satisfied and the total area of the sphere. The surface area where the condition is satisfied is contiguous and confined within the range of $\theta$ and $\phi$, which can be obtained from equation~\ref{lineofsightvelocity} as
\begin{eqnarray}
\theta &=& [0 \ , \ A]  = \left[0 \ , \ \arccos\left(\frac{\zeta}{\sin\phi} \right)\right]\; {\rm and}  \nonumber \\ 
\phi &=& [C \ , \ \frac{\pi}{2}] = \left[\arcsin\zeta \ , \ \frac{\pi}{2}\right] ,
\label{eq_angles1}
\end{eqnarray}
where $A$ and $C$ are the parameters corresponding to the expressions defined in the second square bracket, used as integration limits in equation~\ref{eq_prob}, and $\zeta \equiv (1+q)\, {\tilde a}^{1/2} \,V_{\rm lim}/c$ is a dimensionless parameter.  The probability for detection of the velocity offset is then:
\begin{equation}
P_V=\frac{2}{\pi}\int_{C}^{\pi/2} d\phi \int_{0}^{A} \sin\theta \ d\theta \;.
\label{eq_prob}
\end{equation}
We express $\theta = \theta(\phi)$, since this form simplifies the integral considerably and use symmetry to integrate over one octant and multiply the result by 8. The analytic solution of this integral is 
\begin{equation}
P_V=  1-\frac{2}{\pi}\left[ \arcsin\zeta+\zeta\ln\left( \frac{1+\cos(\arcsin\zeta)}{\zeta}  \right)  \right] \;.
\end{equation}
%

\begin{figure}[t]
\begin{center}
\includegraphics[trim=0 0 0 15, clip, scale=1.0,angle=0]{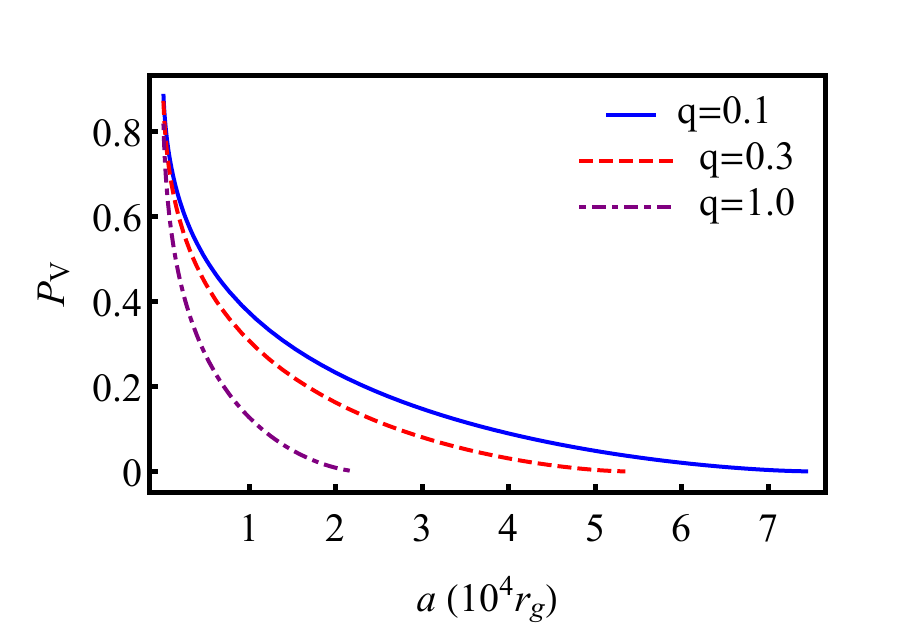} 
\end{center}
\caption{Probability of observing an SBHB of any mass with a velocity offset above the cutoff of $V_{\rm lim} = 1000\, {\rm km\,s^{-1}}$. Different lines mark different mass ratios: 0.1 (blue, solid), 0.3 (red, dashed), 1.0 (purple, dash-dot). SBHBs with larger $a$ and $q$ are less likely to be detected.}
\label{prob_line_sight}
\end{figure}

Figure \ref{prob_line_sight} shows this probability distribution as a function of $a$ for different choices of $q$. As expected, the probability is lower for SBHBs with larger orbital separations, since the velocity is inversely proportional to $a^{1/2}$. $P_V$ also monotonically decreases with $q$, because the line of sight velocity of the secondary SBH is inversely proportional to $(1+q)$. Combined with the orbital orientation effects this property implies that among SBHBs selected for spectroscopic followup\footnote{The statement applies to SBHBs selected based on the initial velocity offset of their broad optical emission lines.} there should be four times as many pairs with $q=0.1$ relative to $q=1$ at orbital separation of $\sim 10^4\,r_g$. Similarly, when projection factors are taken into account, SBHBs characterized by $q=0.3$ and $a = 10^4\,r_g$ satisfy the criterion $V_{\rm obs} \geq V_{\rm lim}$ about 30\% of the time, assuming an isotropic orientation of their orbits.


\subsection{Probability of observing a binary with a significant velocity modulation: $P_{\Delta V}$}
\label{probvelmod}

The criterion used by {\it all} spectroscopic searches to select viable SBHB candidates, regardless of their assumptions about $V_{\rm lim}$, is that the epoch-to-epoch modulation in the observed line of sight velocity must be different from zero and consistent with the SBHB orbital motion. In practice, this implies $\Delta V_{\rm obs} \geq \Delta V_{\rm lim}$ in order for the SBHB to be detected, where 
\begin{equation}
\begin{split}
\Delta V_{\rm obs} &=V_2'\sin(\phi+\Omega \Delta t)\cos\theta-V_2\sin\phi\cos\theta \\
& \approx \frac{1}{1+q}\frac{c}{\tilde{a}^{1/2}}\cos\phi\cos\theta\sin(\Omega \Delta t)\;.
\label{eq:dvobs}
\end{split}
\end{equation}
Here $V_2$ and $V_2'$ are the orbital velocities of the secondary SBH in the earlier and later epoch of observation, respectively, and $\Omega =c^3/(GM\tilde{a}^{3/2})$ is the angular velocity of the SBHB on a circular orbit. $\Delta t$ is the time elapsed between the two measurements of the velocity offset, defined in the frame of the SBHB, which for the E12 search corresponds to time scales from weeks to years. We adopt $\Delta t = 1\,$yr hereafter as a representative value.

Since most SBHBs targeted by the E12 search are expected to have longer orbital periods than the observational baseline, we assume that they traverse only a small portion of their orbit on this time scale. Consequently, $V_2 \approx V_2'$ and $\Omega \Delta t \ll 1$, leading to the approximation in equation~\ref{eq:dvobs} \citep[see also][for a similar approach]{Ju2013}. This assumption in our model breaks down when $\Omega \Delta t \sim 1$ and the time between subsequent observations becomes comparable to the orbital period of the SBHB. We calculate the orbital separations at which this happens for different configurations of SBHBs and list them in Table~\ref{table:cutoff} as $a_{\Delta V}^{\rm min}$. Note that this is merely a limitation of the model presented here, which does not preclude the use of velocity modulation of the broad lines as a criterion for selection of SBHBs with $a < a_{\Delta V}^{\rm min}$ in observations.

From equation~\ref{eq:dvobs} we find the maximum orbital separation of the SBHB associated with the measured $\Delta V_{\rm obs}$ as
\begin{equation}
a_{\Delta V}^{\rm max}=\frac{2.2\times10^4\,r_g}{\sqrt{(1+q)}}
\left(\frac{\Delta V_{obs}}{40\,{\rm km\,s^{-1}}} \right)^{-1/2} 
\left(\frac{\Delta t}{1\,{\rm yr}} \right)^{1/2} 
M_8^{-1/2} ,
\label{eq_a_dV_max}
\end{equation}
where we have taken $\sin(\Omega \Delta t)\approx\Omega \Delta t$, which is justified in the previous paragraph. Note also that for the purposes of deriving equation~\ref{eq_a_dV_max} we have adopted the fiducial value $\Delta V_{\rm obs} = \Delta V_{\rm lim} = 40\, {\rm km\,s^{-1}}$. It follows that all SBHB candidates selected by this survey have orbital semi-major axes smaller than $a_{\Delta V}^{\rm max}$. For illustration we list the values of $a_{\Delta V}^{\rm max}$ corresponding to different SBHB configurations in Table~\ref{table:cutoff}. If the spectroscopic surveys provide measurements of the radial velocity of the primary SBH instead of the secondary, the value of $a_{\Delta V}^{\rm max}$ would be smaller by a factor of $q^{1/2}$.

The spectroscopic searches that impose the selection cutoff in terms of both $V_{\rm lim}$ and $\Delta V_{\rm lim}$ have two upper limits on the maximum value of the SBHB orbital separation, corresponding to $a_V^{\rm max}$ and $a_{\Delta V}^{\rm max}$. In this case, the true detection limit is given by the smaller of the two values. Based on this limit, our model implies that assuming a cadence of observations comparable to $\Delta t = 1\,$yr, and the selection criteria adopted in this model, SBHBs with orbital separations $< {\rm few}\times 10^4\,r_g$ are in principle detectable by spectroscopic searches. If we integrate the probability density $\rho(a)$, given in equation~\ref{eq_rhoa}, from $a_{\rm min} = 10^2\,r_g$ to this cutoff, we find that the SBHBs of interest only reside in this range of separations for $\leq0.5 \%$ of their lifetime. It follows that for every one detected there should be $>200$ undetected gravitationally bound SBHBs with similar properties at larger separations, once the effects of the SBHB orbital orientation are taken into account in addition to their rate of evolution.

\begin{figure}[t]
\begin{center}
\includegraphics[scale=1.0]{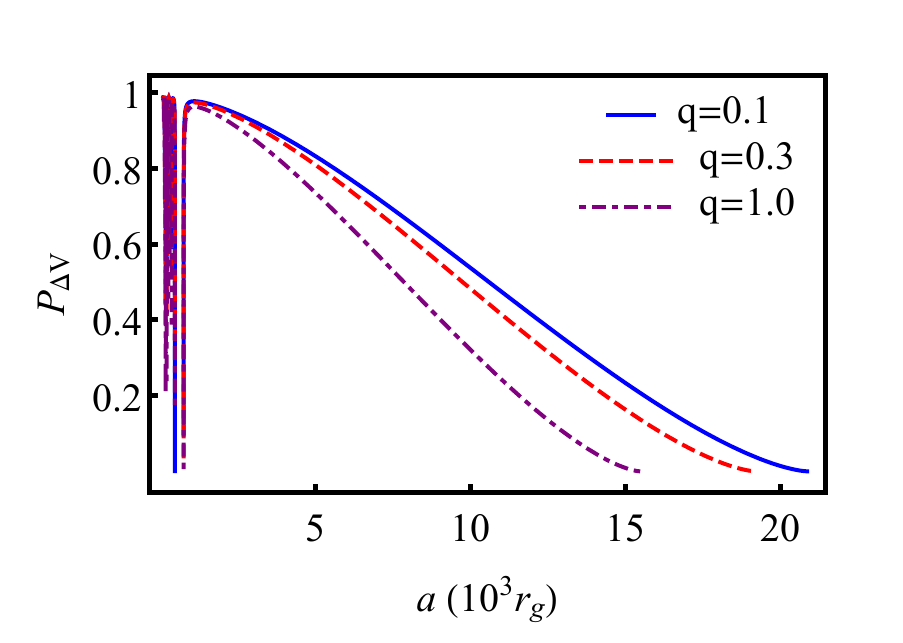} 
\end{center}
\caption{Probability of observing an SBHB with $M=10^8M_\odot$, characterized by the velocity modulation detectable by the E12 spectroscopic search. Different lines mark different mass ratios: 0.1 (blue, solid), 0.3 (purple, dashed), and 1.0 (red, dot-dash). SBHBs with larger $a$ and $q$ are less likely to be detected. The probability drops to zero at small $a$ when the temporal baseline of observations (here assumed to be $\Delta t = 1\,$yr) is comparable to the orbital period of the binary.}
\label{prob_line_mod}
\end{figure}

Solving for the limits on $\theta$ and $\phi$, for which the criterion $\Delta V_{\rm obs} \geq \Delta V_{\rm lim}$ is satisfied, we obtain 
\begin{eqnarray}
\theta&=& [0, B] = \left[ 0 \ , \  \arccos\left(\frac{\xi}{\cos\phi} \right)\right] \nonumber \\ 
\phi&= & [0, D] = \left[0 \ , \ \arccos\xi \right] 
\label{eq_angles2}
\end{eqnarray}
where again we define $\theta$ as $\theta(\phi)$, which allows for exact analytic integration of $P_{\Delta V}$. The parameters $B$ and $D$ set the integration limits for $P_{\Delta V}$ and $\xi$ is the dimensionless parameter defined as
\begin{equation}
\xi\equiv  \frac{(1+q)\,\tilde{a}^{1/2}}{\sin(\Omega \Delta t)}\,\frac{\Delta V_{\rm lim}}{c} 
\label{eq_xi}
\end{equation}
The resulting probability of detection of an SBHB with a given velocity modulation can be defined in the same way as in equation~\ref{eq_prob}. The analytic solution of this integral is 
\begin{equation}
P_{\Delta V}=  \frac{2}{\pi} \left[ \arccos\xi  -\xi \ln\left( \frac{1+\sin\left( \arccos\xi \right)}{\xi} \right) \right] .
\label{eq_PdV}
\end{equation}
Note that $P_{\Delta V}$ is derived in an approximate way in \cite{Ju2013} and that the two solutions differ (see Appendix~\ref{app:Pdv} for comparison).  

Figure \ref{prob_line_mod} shows the resulting probability density function for $10^8M_\odot$ SBHBs with three different mass ratios. Similarly to $P_{V}$, $P_{\Delta V}$ decreases monotonically with $a$ and $q$. In this case, among SBHBs with detected velocity modulation there should be about 60\% more pairs with $q=0.1$ relative to $q=1$ at orbital separations of $\sim 10^4\,r_g$. Similarly, the projection factors lead to one undetected binary for each one observed for SBHBs characterized by $q=0.3$ and $a = 10^4\,r_g$, assuming an isotropic orientation of their orbits.

The behavior of the two probability distributions differs at the smallest values of $a$, where $P_{\Delta V}$ drops to zero where $\Omega\Delta t$ is an integer multiple of $2\pi$. In other words, if the temporal baseline of observations happens to correspond to an integer number of SBHB orbital periods, the spectroscopic observations will return a null measurement of velocity modulation, resulting in a non-detection of the binary. This consideration only applies to spectroscopic searches based on exactly two observations, or multiple, {\it evenly spaced} observations. We elaborate on the calculation of $P_{\Delta V}$ and provide a test of our analytic solution in Appendix~\ref{app:Pdv}.

\begin{figure}[t]
\begin{center}
\includegraphics[trim=0 0 0 15, clip, scale=.6,angle=0]{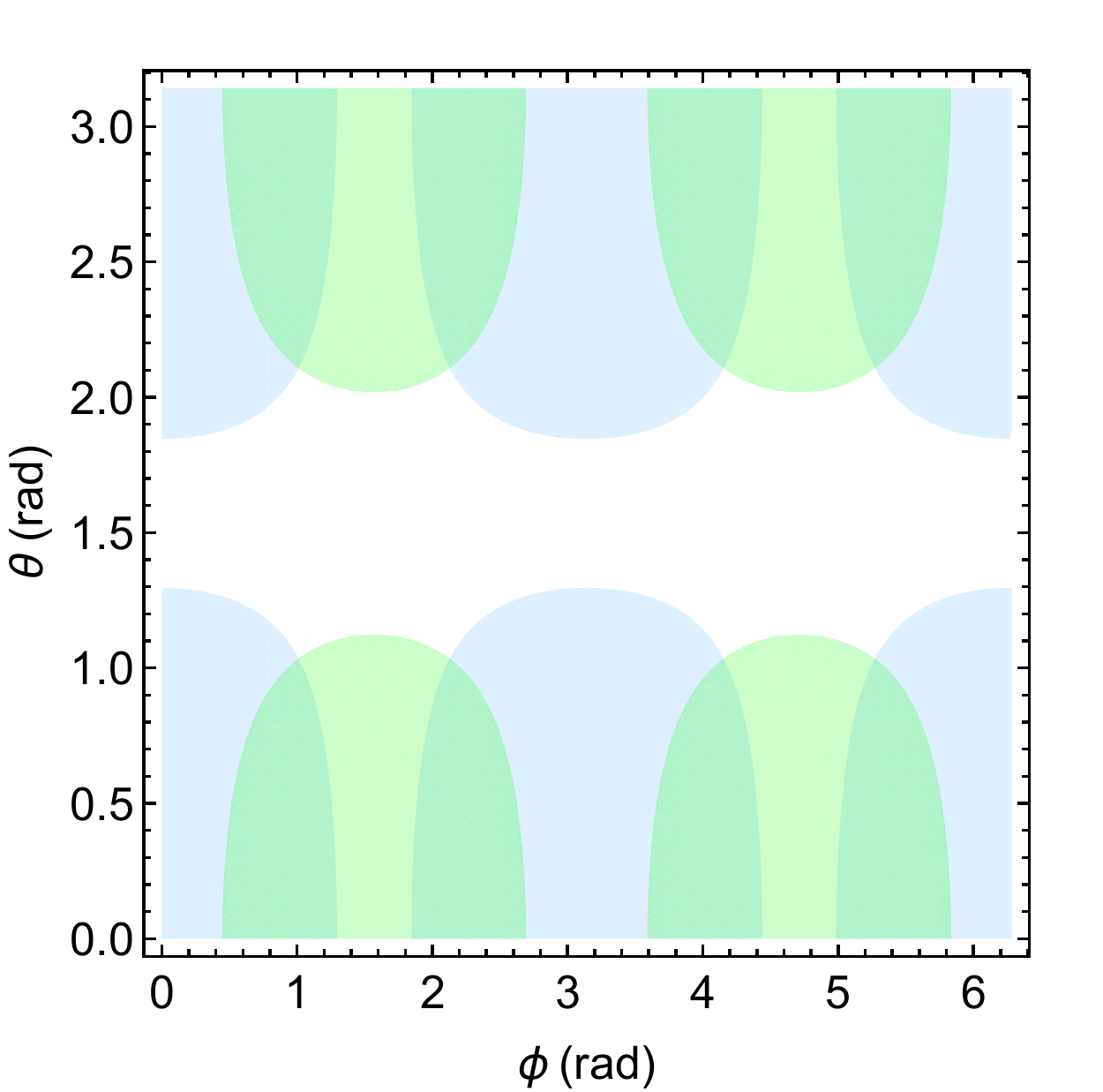} 
\end{center}
\caption{Contour plots of the angles $\theta$ and $\phi$ for which $V_{\rm obs}$ (green) and $\Delta V_{\rm obs}$ (blue) are detectable by the E12 search, calculated for an SBHB with $M=10^8M_\odot$, $q=0.3$, and $a=10^4\,r_g$. The simultaneous detection of both is possible within the regions where the two distributions overlap, corresponding to non-zero $P_{V, \Delta V}$.}
\label{overlap}
\end{figure}


\begin{figure*}[t]
\begin{center}
\includegraphics[trim=0 -5 0 0, clip, scale=0.44,angle=0]{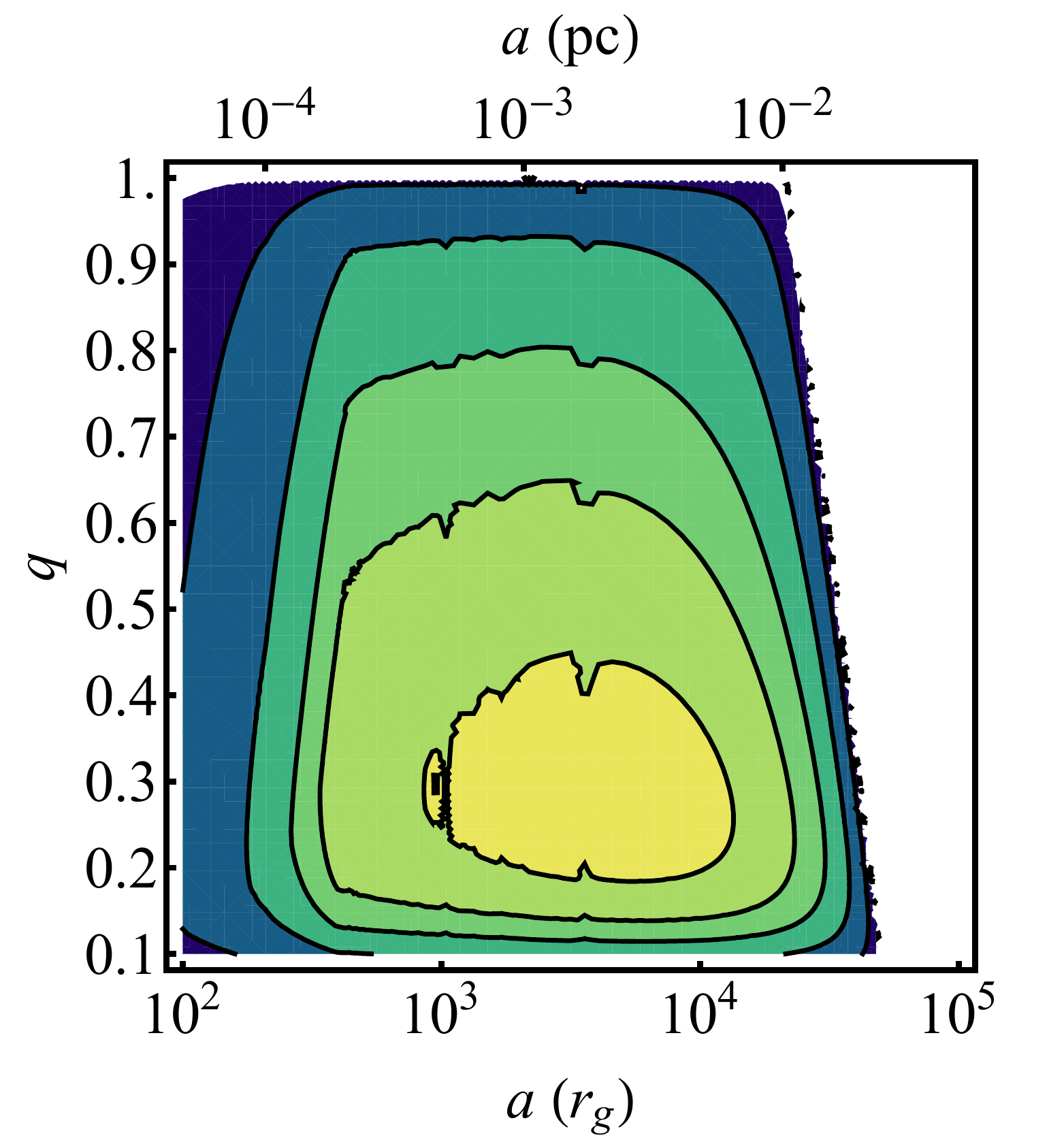} 
\includegraphics[trim=0 0 0 0, clip, scale=0.80,angle=0]{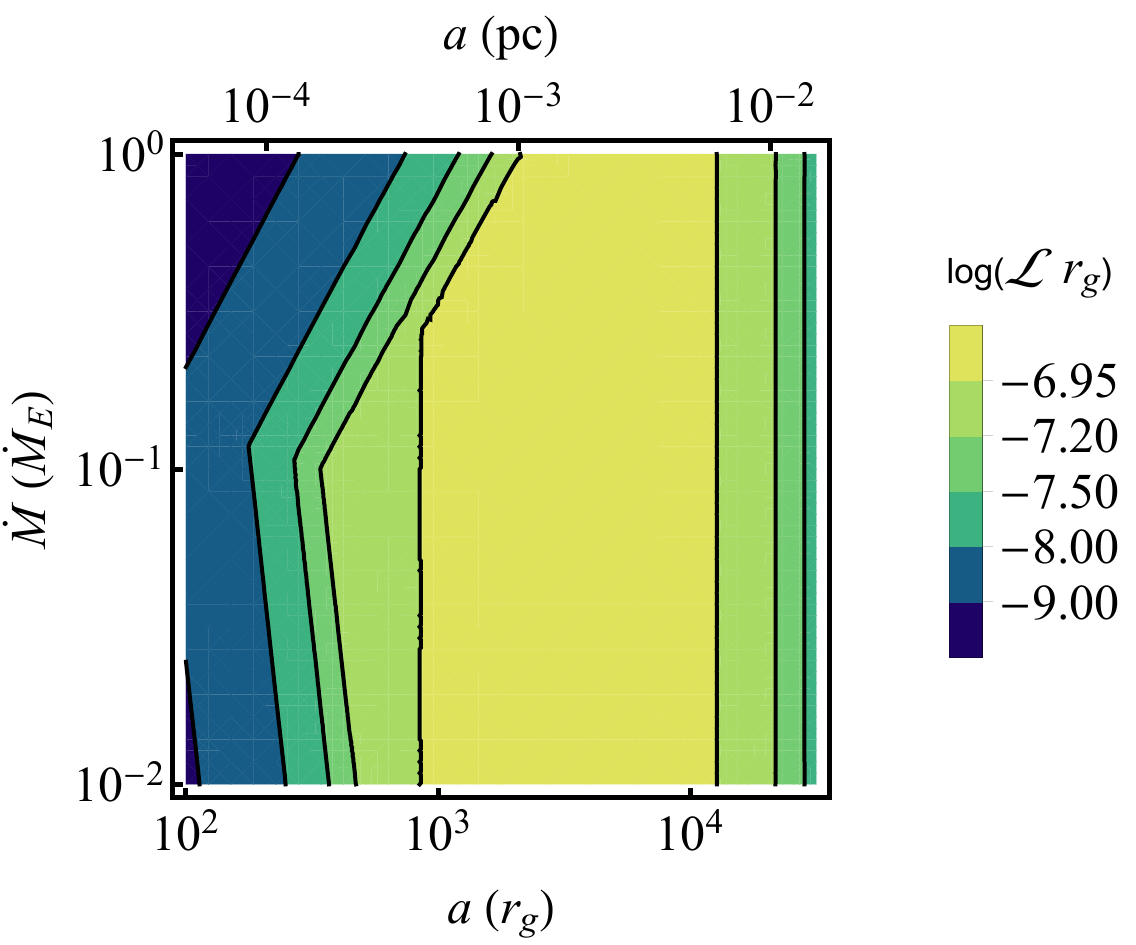} 
\includegraphics[trim=0 -5 0 0, clip, scale=0.44,angle=0]{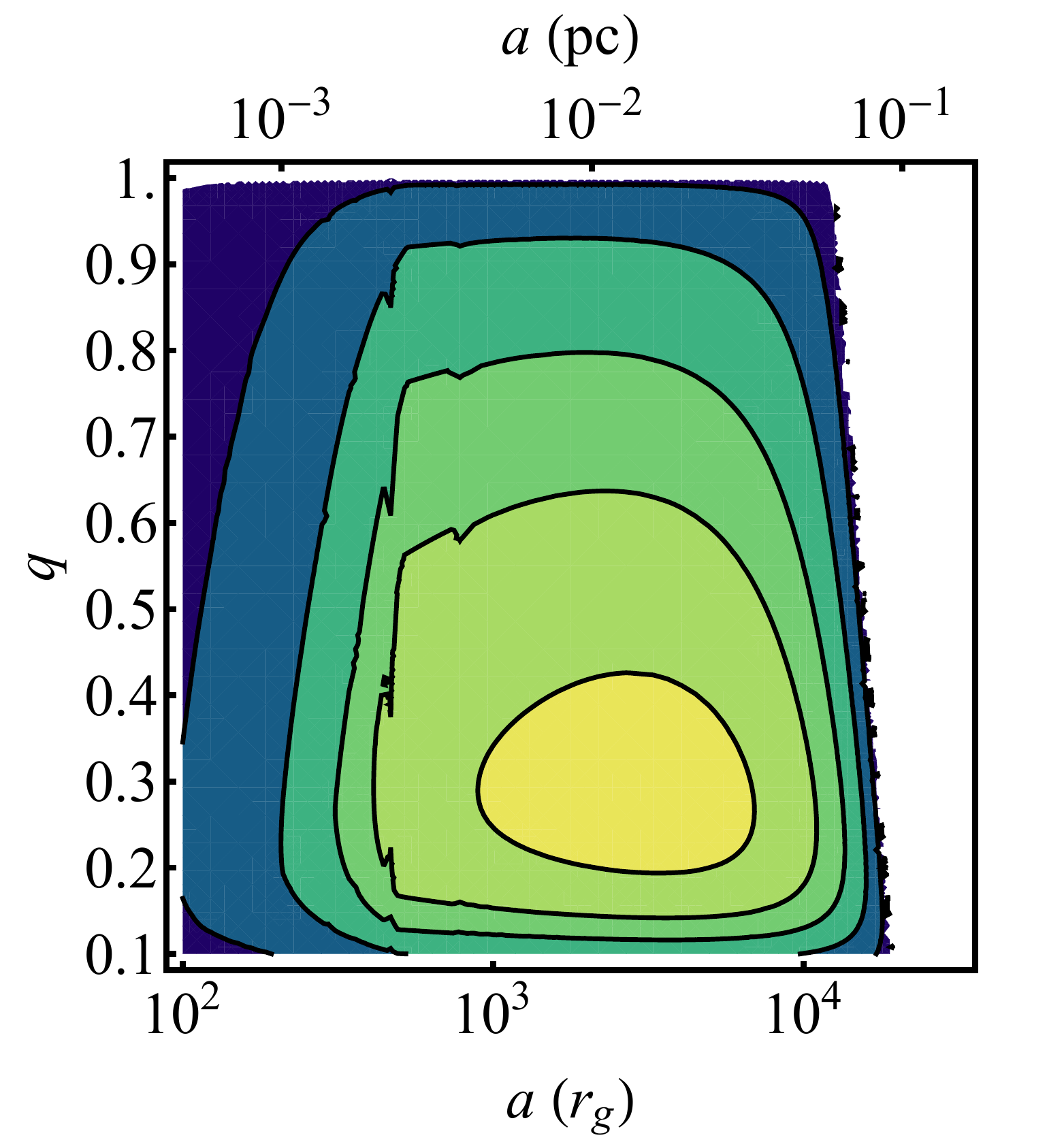} 
\includegraphics[trim=0 0 0 0, clip, scale=0.80,angle=0]{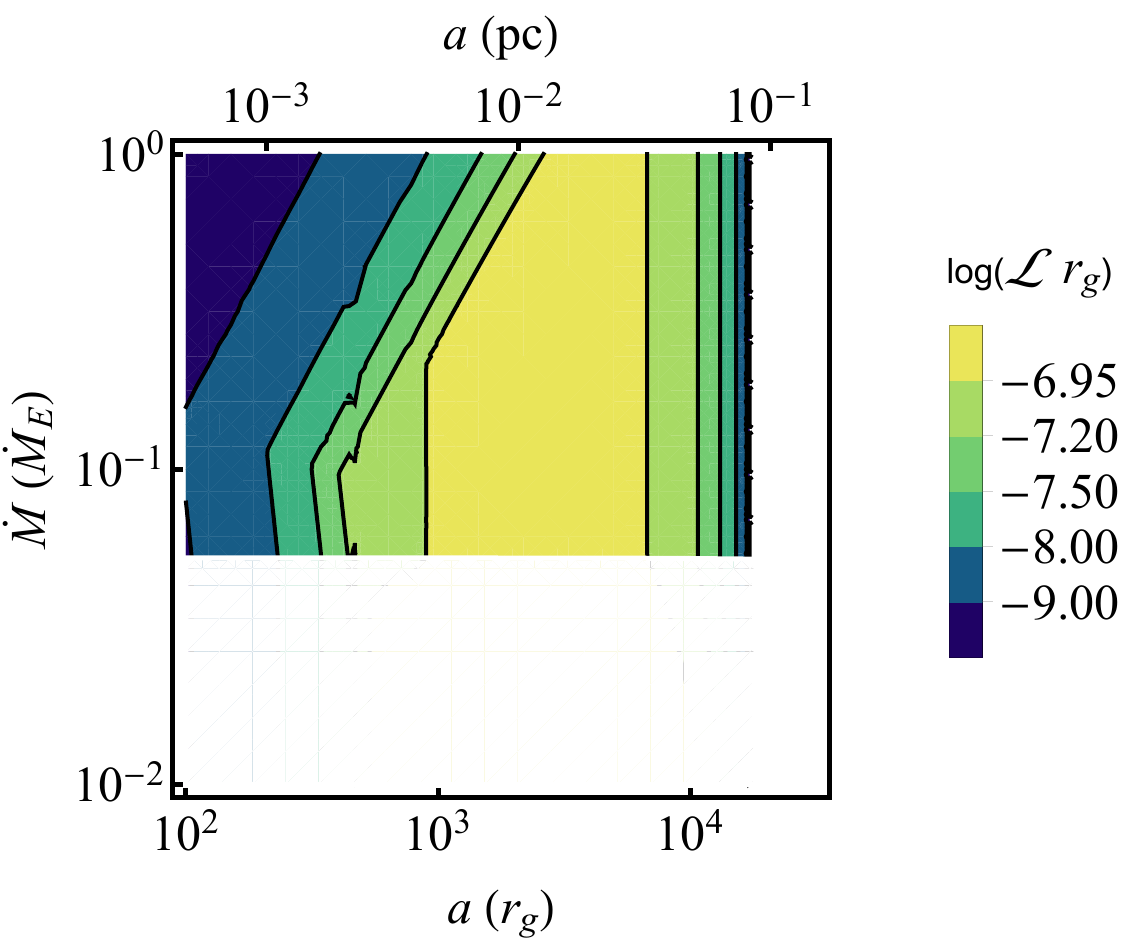} 
\end{center}
\caption{Two-dimensional cuts of the multi-variate likelihood for the E12 search given by $\mathcal{L} = \rho(a)\times \rho(q) \times  P_{V, \Delta V}$. Contour plot of the likelihood distribution for an SBHB with $M=10^7M_\odot$ (top) and $M=10^8M_\odot$ (bottom) for specific choices of $\dot{M}=0.1\dot{M}_E$ (left) and $q=0.3$ (right). White regions mark SBHB configurations where the orbital separations are too wide to be detected by the E12 survey or the orbital evolutionary time scale of the SBHB is longer than a Hubble time. Color marks the logarithmic value of likelihood, $\log(\mathcal{L}\,r_g)$. All panels are plotted on the same color scale and display separate colorbars for easy reading. An open source {\tt python} script for calculation and plotting of the likelihood is available at \url{https://github.com/bbhpsu/Pflueger_etal18}.} 
\label{final_plotz}
\end{figure*}

\begin{figure*}[t]
\begin{center}
\includegraphics[trim=0 0 0 0, clip, scale=0.505]{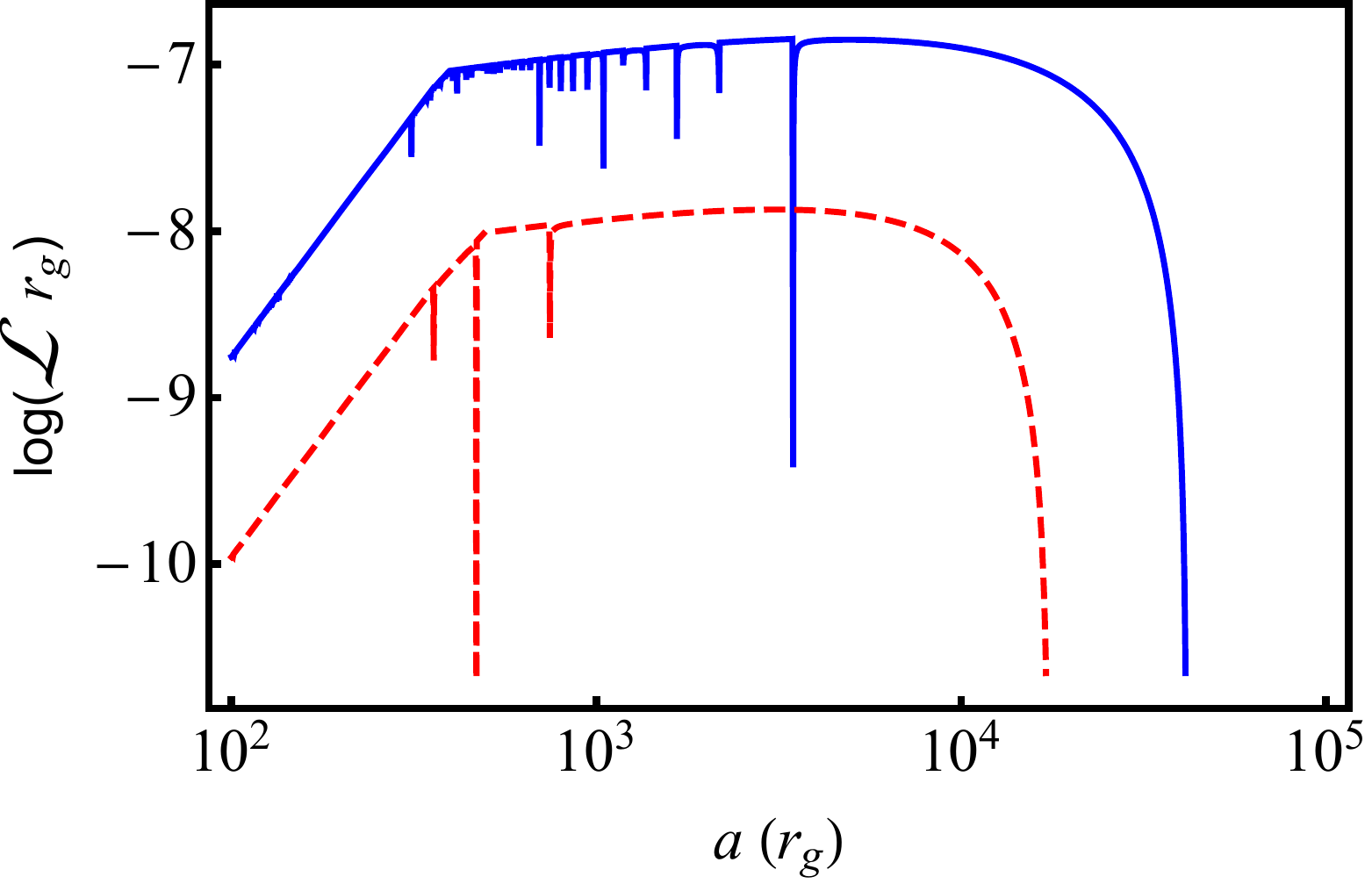} 
\includegraphics[trim=-10 0 0 0, clip, scale=0.5,angle=0]{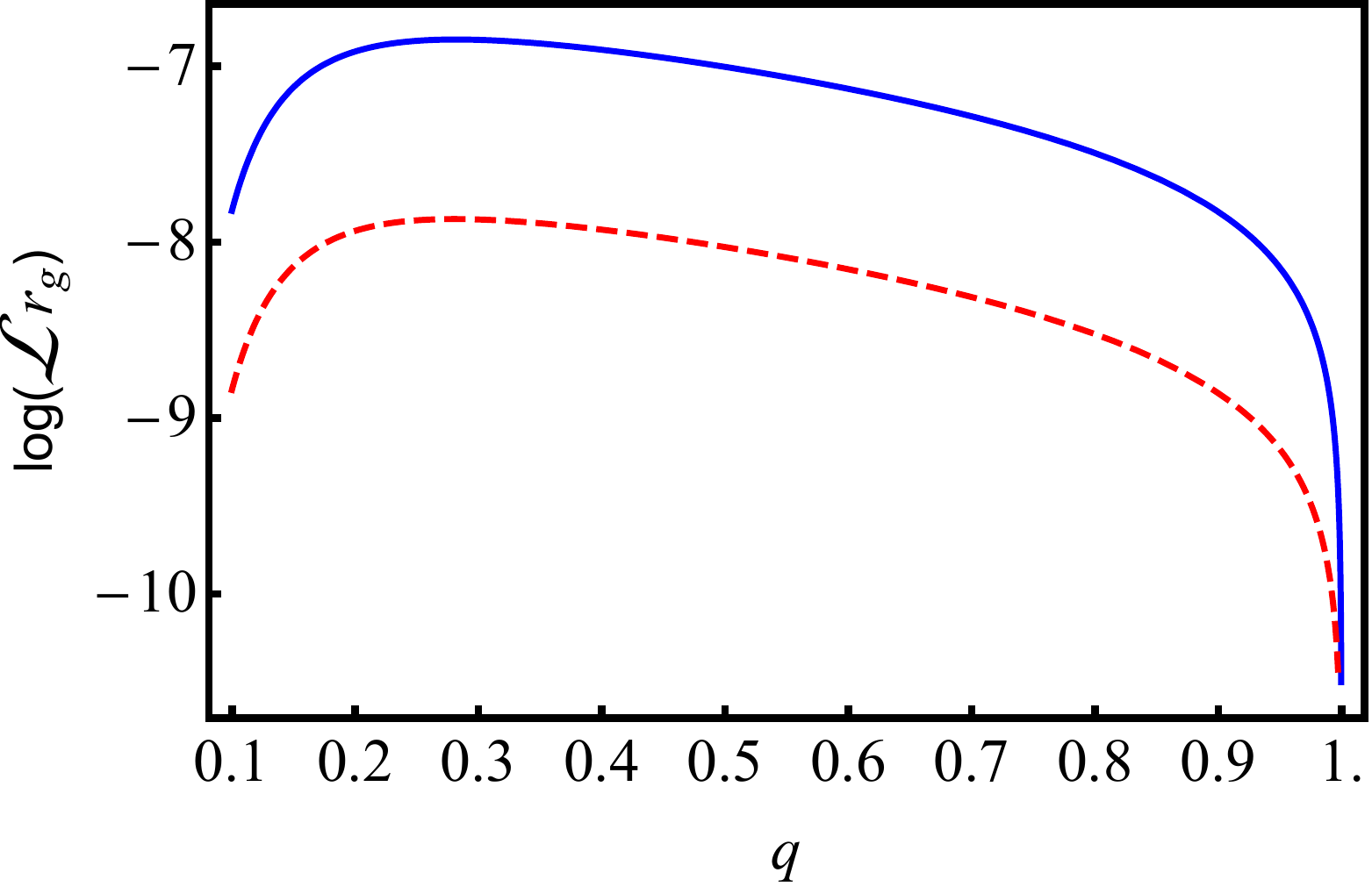} 
\end{center}
\caption{One-dimensional cuts of the multi-variate likelihood, $\mathcal{L}$, corresponding to the two-dimensional distributions in Figure~\ref{final_plotz}, for SBHB mass $10^7M_\odot$ (blue, solid curves) and $10^8M_\odot$ (red, dashed). In both panels we plot $\mathcal{L}/10$ for $10^8M_\odot$ SBHBs for clarity. {\it Left:} $\mathcal{L}$ as a function of $a$ plotted for $q=0.3$. Narrow dips occur at orbital separations where the temporal baseline of observations (here assumed to be $\Delta t = 1$\,year) corresponds to an integer number of SBHB orbital periods. {\it Right:} $\mathcal{L}$ as a function of $q$ shown for $\log\tilde{a}=3.5$.}
\label{fig_1d}
\end{figure*}

\subsection{Probability of simultaneous measurement of $V_{\rm obs}$ and $\Delta V_{\rm obs}$: $P_{V, \Delta V}$}
\label{sec_Pvdv}

The spectroscopic searches that only consider the criterion $\Delta V_{\rm obs} \geq \Delta V_{\rm lim}$ for selection of SBHB candidates should use $P_{\Delta V}$ as the relevant probability distribution. The spectroscopic searches that additionally impose $V_{\rm obs} \geq V_{\rm lim}$ are looking for SBHB candidates satisfying both criteria {\it simultaneously}. In practice this means that, for every plausible SBHB configuration we must find the range of angles of $\theta$ and $\phi$ on the sky for which both criteria are satisfied. Figure~\ref{overlap} illustrates the range of $\theta$ and $\phi$ where each criterion is satisfied as well as the overlap of the two regions for an SBHB with $M=10^8M_\odot$, $q=0.3$, and $a=10^4\,r_g$. The fraction of the solid angle occupied by the overlap is about 17\% indicating that, all other things being equal, one out of six SBHBs with this configuration would be detected, assuming an isotropic orientation of their orbits and that all binaries exhibit AGN-like emission lines.

The overlap can be determined by examining the integration limits for $P_V$ and $P_{\Delta V}$ given in equations~\ref{eq_angles1} and \ref{eq_angles2} for the first octant. The lower limits on $\theta$ are in both cases zero therefore, the lower limit of the $\theta$ integral of $P_{V, \Delta V}$ is zero. The upper limits on $\theta$ for $P_V$ and $P_{\Delta V}$ differ, implying that the overlap is limited by the smaller of $A$ and $B$, or $\min(A,B)$. Examining the limits of $\phi$ shows that the integration lower limit must be $\min(C,D)$ and its upper limit, $\min(D,\pi/2)$.  
\begin{equation}
P_{V, \Delta V}=\frac{2}{\pi}\int_{\min(C,D)}^{\min(D,\pi/2)} d\phi \int_{0}^{\min(A,B)}\sin\theta \; d\theta 
\end{equation}
This results in a probability

\begin{equation}
P_{V, \Delta V}=\frac{2}{\pi}\int_{\min(C,D)}^{\min(D,\pi/2)}\nu(A,B) \; d\phi
\label{Pvdv}
\end{equation}
where we have integrated over $\theta$ and provided a simplified expression in terms of $\phi$ that can be integrated numerically. The function $\nu$ is defined as 
\begin{equation}
\nu(A,B)\equiv\left[1-\frac{\cos A}{2}-\frac{\cos B}{2}- 
 \left|\frac{\cos B}{2}-\frac{cos A}{2}   \right|\right] .
\end{equation}
Note that if $D<C$ the $\phi$ the integral returns a zero probability (i.e., no overlap), since all angles are considered in the first octant and must be $\leq\pi/2$. In Appendix~\ref{app:Pvdv} we show a plot of $P_{V, \Delta V}$, provide calculation details, and a test of our numerical solution.


\subsection{Implications for spectroscopic searches}
\label{sec:implications}

In this section we present the overall likelihood for detection of a sub-parsec SBHB by the E12 spectroscopic search by combining the intrinsic probability distributions with the observational selection effects. The likelihood in this case is defined by the product $\mathcal{L} \equiv \rho(a)\times \rho(q) \times  P_{V, \Delta V}$\footnote{In the case of spectroscopic searches that do not impose a cutoff in $V_{\rm lim}$, this likelihood can be calculated as $\rho(a)\times \rho(q) \times  P_{\Delta V}$.}. Therefore, the dependence of $\mathcal{L}$ on the properties of SBHBs and the criteria of observational searches is
\begin{equation}
\mathcal{L}(a, q\,|\,M,\dot{M}, V_{\rm obs} \geq V_{\rm lim}, \Delta V_{\rm obs} \geq \Delta V_{\rm lim})\;.
\label{eq_Ldependence}
\end{equation}
Here $\int da \int dq\, \mathcal{L} = 1$ only when $V_{\rm lim}$ and $\Delta V_{\rm lim}$ are zero (and hence $P_{V, \Delta V} =1$), implying that, in this special case $\mathcal{L}$ becomes a true probability density. In all other cases the value of this integral $<1$ and so we refer to $\mathcal{L}$ as the likelihood, in general\footnote{We provide an open source {\tt python} script for calculation and plotting of the likelihood at \url{https://github.com/bbhpsu/Pflueger_etal18}.}.

It is worth noting that multiplication of the probability densities involves an implicit assumption that the they are independent. This is a reasonable assumption for our model.  As noted in \S~\ref{subsec:orbitevol}, the description of $\rho(a)$ implies that the evolution of a gravitationally bound SBHB has a weak dependence on $q$ (and only for SBHBs in the GW regime). Therefore, $\rho(q)$ is primarily shaped by the processes preceding the formation of a bound SBHB. $P_{\Delta V}$ and $P_{V, \Delta V}$ are the probability distribution functions that account for the projection effects and provide the of SBHBs with a given configuration that can be detected by spectroscopic searches, if their orbits are isotropically oriented in space. Both only modify the intrinsic probability density a posteriori. 

Figure~\ref{final_plotz} shows the resulting multi-variate likelihood for the E12 search, which is a function of $M$, $q$, $a$ and $\dot{M}$. The likelihood is plotted for two choices of SBHB masses, $M = 10^7M_\odot$ (top) and $10^8M_\odot$ (bottom). The two-dimensional cut in the left panels illustrates the likelihood in terms of $a$ and $q$ for an SBHB characterized by $\dot{M}=0.1\dot{M}_E$. The most notable feature of this plot is the ``bull's eye'' region which indicates the range of $a$ and $q$ most likely to be detected by the E12 search. For both SBHB masses this range is $10^3\,r_g \lesssim a \lesssim 10^4\,r_g$, albeit centered on somewhat smaller separations for $10^8M_\odot$ SBHBs. In both cases SBHBs most likely to be detected have $0.2 \lesssim q \lesssim 0.4$. The white region at large $a$ represents the cutoff beyond which SBHBs are not detectable because their values of $V_{\rm obs}$ and $\Delta V_{\rm obs}$ fall below the limits set by the search. We do not show the likelihood for $10^9M_\odot$ SBHB systems but note that, in agreement with the above trend, the peak of their distribution is located at even smaller semi-major axes, when expressed in unites of $r_g$. 

The right panels of Figure~\ref{final_plotz} show a different cut of the same likelihood distribution calculated for a given SBHB mass and $q=0.3$. They show a range of favored SBHB orbital separations similar to the left panels as well as a plausible range of $\dot{M}$ for each system. $10^7M_\odot$ SBHBs are expected to reach the detectable range of $a$ even at lower values of $\dot{M}$, while $10^8M_\odot$ SBHBs are detectable only when $\dot{M} > 0.06\dot{M}_E$. This lower limit corresponds to the requirement that the transport of angular momentum must be efficient enough in order for an SBHB to reach detectable orbital separations in less than a Hubble time. For SBHB systems with a mass of $10^9M_\odot$, the cutoff shifts to higher values still, $\dot{M} \sim \dot{M}_E$, indicating that any mechanism responsible for transport of SBHBs must be at least as effective as a disk with an accretion rate comparable to the Eddington rate. Note that this is a conservative lower limit, as we only require that the evolutionary time scale of a gravitationally bound SBHB is shorter than the age of the universe. We do not account for the time it takes the two galaxies to merge and the SBHs to pair, which by itself may take several gigayears \citep[e.g.,][]{kazantzidis05}.

Figure~\ref{fig_1d} shows one-dimensional cuts of the multi-variate likelihood, $\mathcal{L}$, corresponding to the two-dimensional distributions in Figure~\ref{final_plotz} for SBHBs with mass $10^7M_\odot$ and $10^8M_\odot$. In both panels, the likelihood plotted for $10^8M_\odot$ SBHBs is divided by a factor of 10, for clarity of presentation. The left panel illustrates that the likelihood distribution has similar shape for both SBHB masses but that the more massive SBHBs have a somewhat narrower range of orbital separations within which they can be detected by spectroscopic searches. Similarly, the two curves exhibit narrow dips in the likelihood at orbital separations where the temporal baseline of observations corresponds to an integer number of SBHB orbital periods, resulting in a non-detection of the binary. These dips in one-dimensional plots directly correspond to the ``notches" in contours noticeable in the two-dimensional maps of Figure~\ref{final_plotz}.

The right panel of Figure~\ref{fig_1d} shows relatively broad and flat likelihood distribution for both $10^7M_\odot$ and $10^8M_\odot$ SBHBs, which peaks at $q\approx 0.3$. The location of the peak reflects the assumed distribution of SBHB mass ratios, shown in Figure~\ref{massratiodis}, and may change if a different description of $\rho(q)$ is adopted.


\section{Discussion}
\label{sec:discussion}


\subsection{Benefits of long-term monitoring and complementary searches for SBHBs}
\label{sec:benefits}

The likelihood distribution presented in the previous section can, in principle, be calculated for every single SBHB candidate, given a minimum of two spectroscopic measurements of velocity offset, necessary to establish $V_{\rm obs}$ and $\Delta V_{\rm obs}$. The true power of SBHB monitoring however comes from repeated measurements, which define the velocity curve of the SBHB with increasing confidence. More specifically, repeated measurements provide constraints on $V_2 \cos\theta$, the amplitude of the velocity curve and the orbital period $P$. Because $V_2 \cos\theta  = V_{\rm obs} / \sin\phi \geq V_{\rm obs}$, the amplitude provides a stronger constraint on the orbital separation than just two measurements of the velocity offset. This constraint would be represented by the following curve in the likelihood maps of Figure~\ref{final_plotz}
\begin{equation}
\tilde{a} <  (1+q)^{-2} \left(\frac{V_2\cos\theta}{c} \right)^{-2} \;.
\end{equation}

Note that the spectroscopic measurements alone cannot uniquely confirm the identity of an SBHB, because for the majority of configurations considered in this model the binary orbital period is longer than the temporal baseline of observations. This precludes measurement of multiple orbital cycles, which has traditionally been used as a criterion for binarity in stellar systems. Therefore, in the case of SBHB candidates the spectroscopic observations need to be  supplemented by additional, independent observational evidence that can help to further elucidate their nature. 

Some of the more promising complementary approaches include the direct VLBI imaging at millimeter and radio wavelengths of nearby SBHB candidates ($z\lesssim 0.1$) with separations $a \gtrsim 0.01\,$pc \citep{dorazio17}. The two approaches are complementary since VLBI is not a surveying technique and it depends on other approaches to define the candidate sample. The emission properties of SBHB candidates at these wavelengths are however largely unknown and difficult to uniquely predict from theory, making such observations risky but potentially highly rewarding, should any SBHBs be detected.

It is worth noting that SBHBs that are in principle accessible to spectroscopic surveys also overlap with a  population targeted by the ongoing photometric surveys and PTAs. These are the SBHBs with orbital periods of a few years 
\begin{equation}
P = 3.1\,{\rm yr} \left(\frac{a}{10^3\,r_g} \right)^{3/2} M_8 
\end{equation}
corresponding to the more compact systems ($a \lesssim 0.01\,$pc) considered in this model. The possibility of detection of SBHB candidates with more than one technique is exciting as it may provide additional means to test their nature.

Note that candidates with velocity curves which are inconsistent with the SBHB model can be ruled out based on spectroscopic observations, even if they have been monitored for less than a full orbital cycle. Hence, the spectroscopic followup alone can be effective in narrowing down the sample of SBHB candidates by rejecting those inconsistent with the binarity hypothesis. This approach is adopted and laid out in \citet{Runnoe2017}.


\subsection{Simplifying assumptions and their implications}
\label{S_simplifications}

One important assumption made in this work is that $M$, $q$, and $\dot{M}$ are constant throughout the SBHB evolution and that the binary orbit remains circular. While the evolution in time of any of these parameters is fairly uncertain, there is general understanding of how they may affect the binary orbit. For example, the SBHB may increase its mass though accretion of gas, which is assumed to happen for at least a fraction of SBHB evolutionary time, when the SBHB is detectable though emission of broad lines. 

Similarly, simulations of SBHBs in circumbinary disks find that the secondary SBH tends to accrete at a higher rate than the primary \citep{al96, hayasaki07, roedig11, farris14}. If so, the early evolution of the SBH pairs towards more equal-mass ratios, discussed in \S~\ref{S_params}, is expected to continue after the gravitationally bound SBHB has formed. This is, for example, indicated by the Illustris simulation in which $q \approx 1$ pairs dominate over all other mass ratios at SBH separations of $\sim 1\,$kpc \citep[see Figure~2 in][]{Kelley2017}. If this mode of growth is favored, it is plausible that by the time they reach orbital separations accessible to the spectroscopic searches ($a \lesssim {\rm few}\times 10^4\,r_g$) the SBHB may already be quite close to $q \approx 1$.  The increase in $M$ and $q$ can lead to the shrinking of the SBHB orbit as a consequence of conservation of orbital angular momentum, and so can evolution from an initially eccentric to a circular orbit. It is however expected that these processes are secondary to the circumbinary disk torques in driving the orbital evolution \citep[see][for a comprehensive review of this issue]{rafikov16}, which is in our model encoded in $\dot{M}$.

It is worth emphasizing that in this model $\dot{M}$ is used as a free parameter that describes an effective rate of orbital evolution that can arise due to the circumbinary disk {\it combined} with additional mechanisms unrelated to accretion torques. As such, $\dot{M}$ sets the upper limit on but does not imply the mass accretion rate onto the two SBHs throughout their orbital evolution\footnote{For this reason we refrain from making predictions about the luminosity or the growth in mass of the SBHs based on $\dot{M}$.}. The SBHBs can also be transported to sub-parsec scales through (scattering) interactions with stars \citep{berczik06, preto11, khan11,khan13}. Those SBHBs that reside in clumpy disks may undergo multiple scatterings resulting in stochastic orbital evolution \citep{fiacconi13, roskar14}. These processes are not explicitly captured in this analytic model, which assumes that $\dot{M}$ does not change in time or as a function of the orbital separation of a binary. It is, nevertheless, expected that once a circumbinary disk is in place, the evolution of the SBHB due to its interaction with this disk should proceed efficiently, on a time scale shorter than that for evolution by gravitational scattering. This can be ascribed to the ``conveyor belt" nature of the disk which transports the SBHB angular momentum to large distances, where it takes only a small amount of mass to absorb it.

It is worth noting that the rate of binary evolution is expected to decrease in the {\it secondary-dominated regime}: the stage in which the mass of the secondary SBH dominates over the local circumbinary disk, and the circumbinary disk is the sole driver of orbital evolution \citep{haiman09, Rafikov2013}. In this scenario, the disk cannot effectively ``absorb" the SBHB's orbital angular momentum leading to longer residence times of the binary. We do not account for this effect and as a consequence possibly overestimate the rate of evolution of SBHBs in this regime. For example, SBHBs with $M>10^7\,M_\odot$ and $q>0.1$ transition into the secondary-dominated regime when they are within the detection window of spectroscopic searches, at separations $\lesssim {\rm few}\times 10^4\,r_g$ \citep[see Figures 3 and 4 in][]{haiman09}. The longer residence-time of SBHBs in this range would lead to an increased likelihood of their detection, relative to predictions reported in this work.

In this work we only consider SBHBs on circular orbits for simplicity. Allowing for eccentric orbits would primarily result in larger maximum orbital separations given by $a_V^{\rm max}$ and $a_{\Delta V}^{\rm max}$, because eccentric binaries sweep over a wider range of speeds and accelerations over the course of one orbit, relative to their circular counterparts. Therefore, eccentric SBHBs that reach the detection window of spectroscopic searches would be characterized by a wider range of orbital separations. Their probability and likelihood distributions would nevertheless remain qualitatively similar to those reported in this work. Taking into consideration that addition of one more free parameter (eccentricity) would result in a considerably more complicated model, we choose a simpler (circular) model without much loss of generality.

An implicit assumption made in our model is that the optical BLR of the secondary SBH is present and its signatures detectable over a range of relevant SBHB separations that span from $10^2\,r_g$ to ${\rm few}\times 10^4\,r_g$. In reality, the extent and emission properties of the BLRs surrounding SBHs in a binary are largely unknown. Even without tangible proof, their existence seems plausible in at least a fraction of SBHBs, because they are so ubiquitous in regular AGN. This question can be directly tested by a combination of observational techniques described in \S~\ref{sec:benefits} once a robust sample of SBHBs is established. For example, if the spectroscopic surveys show a lack of SBHBs, relative to the numbers detected by the photometric surveys or inferred from the PTA measurements, this would point to phenomenology different from BLRs of regular AGN.

Finally, in this model we make no assumptions about the underlying SBHB mass function (instead, $M$ is treated as a free parameter) and luminosity distribution function, because they are highly uncertain. Should these properties be better constrained in the future, they can be added to the model a posteriori.


\section{Conclusions}
\label{sec:conclusions}

We presented an analytic model that can be used to determine the likelihood of detection of sub-parsec SBHBs in current and future spectroscopic searches. The model uses a simple theoretical prescription for orbital evolution of SBHBs in circumbinary disks, from an instance when they form a gravitationally bound pair to the point of coalescence. Combined with the selection effects of spectroscopic surveys, it returns a multivariate likelihood for SBHB detection as a function of $M$, $q$, $a$ and $\dot{M}$. 

Our model indicates that a fraction of SBHBs should remain undetected because their evolution from larger separations and into the detection window of the spectroscopic surveys exceeds a Hubble time. For an SBHB to have a chance of being detected, the transport of angular momentum must be faster than that of a disk with mass accretion rate $\sim 10^{-2}\,\dot{M}_E$ for binaries with mass $10^7M_\odot$, whereas this threshold increases to $\sim \dot{M}_E$ for $10^9M_\odot$ systems. This confirms findings of some earlier studies that orbital evolution of the most massive SBHBs (and in particular those that form at $z<2$) may need a helping hand from additional mechanisms, besides circumbinary disks, in order to make it to sub-parsec scales and coalescence \citep{dotti15}.

We find that most SBHBs, which are in principle detectable by spectroscopic surveys with yearly cadence of observations, are expected to reside at orbital separations $\lesssim {\rm few}\times 10^4\,r_g$, where they spend $\leq 0.5 \%$ of their life as gravitationally bound binaries.  Therefore, for every one system in the detectable range of orbital separations there should be $\geq 200$ undetectable gravitationally bound SBHBs with similar properties, at larger separations. Of those that already occupy the right range of separations, the effects of orbital orientation will allow detection of only one in six SBHBs characterized by $M=10^8M_\odot$, $q=0.3$, $a=10^4\,r_g$ and an isotropic orientation of orbits. The latter statement applies if spectra of all SBHBs in this range exhibit the AGN-like emission lines utilized by spectroscopic searches, implying an even larger number of undetected binaries if some fraction of them is inactive.

The surveys should be sensitive to SBHBs with comparable mass ratios, which are generically predicted by theoretical models of pairing and evolution of binaries in gaseous media. Our model suggests that among the SBHBs selected for spectroscopic followup there should be four times as many pairs with $q = 0.1$ relative to $q = 1$ at orbital separations of $\sim10^4\,r_g$. 

Finally, this approach already allows us to calculate the most likely orbital parameters for individual spectroscopic SBHB candidates and to constrain their rate of orbital evolution, if they are genuine binaries. The true power of spectroscopic (and other) searches however comes from continued monitoring, which should lead to improved inferences of binary parameters, especially for systems with shorter orbital periods comparable to the baseline of observations. If such SBHB systems exist, they can be targeted with both spectroscopic and photometric searches, in combination with PTA measurements of the stochastic GW background. The nearby systems ($z < 0.1$) with longer orbital periods ($\gtrsim 10^2\,$yr) are a good match for high angular resolution of VLBI, which can be used to directly image the emission from the radio ``cores'' powered by the two SBHs.

\acknowledgments 
The authors would like to thank Kirk Barrow, Massimo Dotti, and Luke Kelley for insightful discussions that helped to advance this work. T.B, B.P., and K.N. acknowledge support by the National Aeronautics and Space Administration under Grant No. NNX15AK84G issued through the Astrophysics Theory Program and by the Research Corporation for Science Advancement through a Cottrell Scholar Award. T.B., M.E., J.R., and S.S. acknowledge the support from grant AST-1211756 from the National Science Foundation. One part of this work was performed at the Aspen Center for Physics, which is supported by National Science Foundation grant PHY-1607611. T.B. is a member of the Multiple AGN Activity (MAGNA) project that investigates activity in systems of dual and multiple supermassive black holes (\url{http://www.issibern.ch/teams/agnactivity}).

\appendix
\label{appendix}
\section{A: Characteristic Orbital Separations} \label{app:radii}

Here we describe the calculation of characteristic binary orbital separations which determine the boundaries between the regimes that an SBHB evolves through. These orbital separations represent the limits of the integral in equation~\ref{eq_tx}, used to evaluate the time that SBHB spends in each evolutionary regime. Namely, after the gravitationally bound binary forms at $a_{\rm max} \approx 10^6\,r_g$, it can evolve through the gas pressure dominated, radiation pressure dominated region of the circumbinary disk, and finally, GW regime to coalescence. We find that while some SBHBs can pass through all three regimes, other binaries may transition into the GW phase directly from the gas pressure dominated regime (see Figure~\ref{fig:regimes}). Which scenario plays out depends on the properties of the binary and the circumbinary disk.

We therefore evaluate the characteristic orbital separations for all relevant transitions by equating the infall rates given in equations~\ref{eq_adot_g}--\ref{eq_adot_gw} as follows
\begin{eqnarray}
& \dot{a}_{\rm gas}=\dot{a}_{\rm rad},\;\;\; & a_{\rm ch1}= 3550\,r_g\,\alpha^{2/21}\, \dot{m}^{16/21}\, M_8^{2/21} ,\\
&  \dot{a}_{\rm rad}=\dot{a}_{\rm gw},\;\;\; & a_{\rm ch2}= 1.13\times 10^{-2}\, r_g\,\alpha^{-2}\,\frac{q^2}{(1+q)^4}\,\dot{m}^{-4}, \\
&  \dot{a}_{\rm gas}=\dot{a}_{\rm gw},\;\;\; & a_{\rm ch3} = 311\, r_g\,\alpha^{-8/26}\,
\frac{q^{10/26}}{(1+q)^{10/13}}\, \dot{m}^{-2/13}\, M_8^{1/13}  .
\end{eqnarray}


\section{B: Mass Ratio Probability Distribution}
\label{app:q}

As described in \S~\ref{qprob}, the adopted mass ratio probability distribution is motivated by cosmological simulations that follow mergers of dark matter halos \citep{Stewart2009, Hopkins2010}
\begin{equation}
\rho(q) \propto q^{-0.3}(1-q)
\label{eq_b1}
\end{equation}
Relating this expression to the mass ratio distribution of SBHBs requires an implicit assumption that more massive halos will on average host more massive black holes, as in \citet{Lousto2012}, for example. However, there is a clear expectation, based on the simulations of SBH pairing in galactic mergers, that the distribution for the SBHBs ought to depart from that for the dark matter halos at low values of $q$. Namely, while the above expression diverges as $q$ approaches zero, the simulations of SBH pairing indicate that pairs with mass ratios $q<0.1$ should be rare (see \S~\ref{S_params} in the main text). To address this expectation we modify the mass ratio probability distribution by multiplying the expression in equation~\ref{eq_b1} by a Gaussian in $1/q$ to obtain
\begin{equation}
\rho(q)=\gamma\, q^{-0.3}(1-q)\,e^{-\beta/q^2} ,
\label{b2}   
\end{equation}
where $\beta$ and $\gamma$ are the dimensionless parameters. To normalize the distribution in equation~\ref{b2} we require that
\begin{equation}
\int_0^1 \rho(q) \ dq =
\gamma \left[\int_0^1  q^{-0.3}e^{-\beta/q^2} \ dq  -\int_0^1  q^{0.7}e^{-\beta/q^2} \ dq  \right] = 1
\label{b4}
\end{equation}
With a change of variables, $t=\beta/q^2$ and $dq = -(\beta^{0.5}\,t^{-1.5} dt)/2$, one obtains
\begin{equation}
\int_0^1 \rho(q) \ dq =\frac{\gamma}{2}\left[\beta^{0.35} \int_\beta^\infty t^{-1.35}e^{-t}\  dt - \beta^{0.85} \int_\beta^\infty t^{-1.85}e^{-t}\  dt \right] 
\end{equation}%
The two integrals can be expressed in terms of the ordinary and lower gamma functions, given here for completness 
\begin{equation}
\begin{split}
\Gamma(z)\equiv\int_0^\infty t^{z-1}e^{-t} \ dt   \;\;\; {\rm and} \;\;\; 
\Gamma_{\rm L}(z,x)\equiv\int_0^x t^{z-1}e^{-t} \ dt,  \;\;\;\;\;\;
{\rm where}  \;\;\;\;\;\;
\Gamma_{\rm L}(z,x)=x^z\sum_{j=0}^\infty (-1)^j\frac{x^j}{j!(j+z)}
\end{split}
\label{b3}
\end{equation}
Noting that the gamma function is defined for negative non-integer values, we make use of the series expansion for the lower gamma function \citep{Abramowitz1970,Blahak2010}. It follows that
\begin{equation}
\int_0^1 \rho(q) \ dq = \frac{\gamma}{2}\left[ \beta^{0.35}\left[ \Gamma(-0.35)-\Gamma_{\rm L}(-0.35,\beta)  \right] -\beta^{0.85}\left[ \Gamma(-0.85)-\Gamma_{\rm L}(-0.85,\beta)  \right]  \right] = 1
\end{equation}
Evaluating this expression and imposing the constraint that the mass ratio density has a peak at $q=0.3$ yields $\gamma=2.78907$ and $\beta=0.0327857$. 



\section{C: Comparison of analytic and Monte Carlo solutions for $P_{\Delta V}$}
\label{app:Pdv}

\begin{figure}[t]
\begin{center}
\includegraphics[scale=0.6]{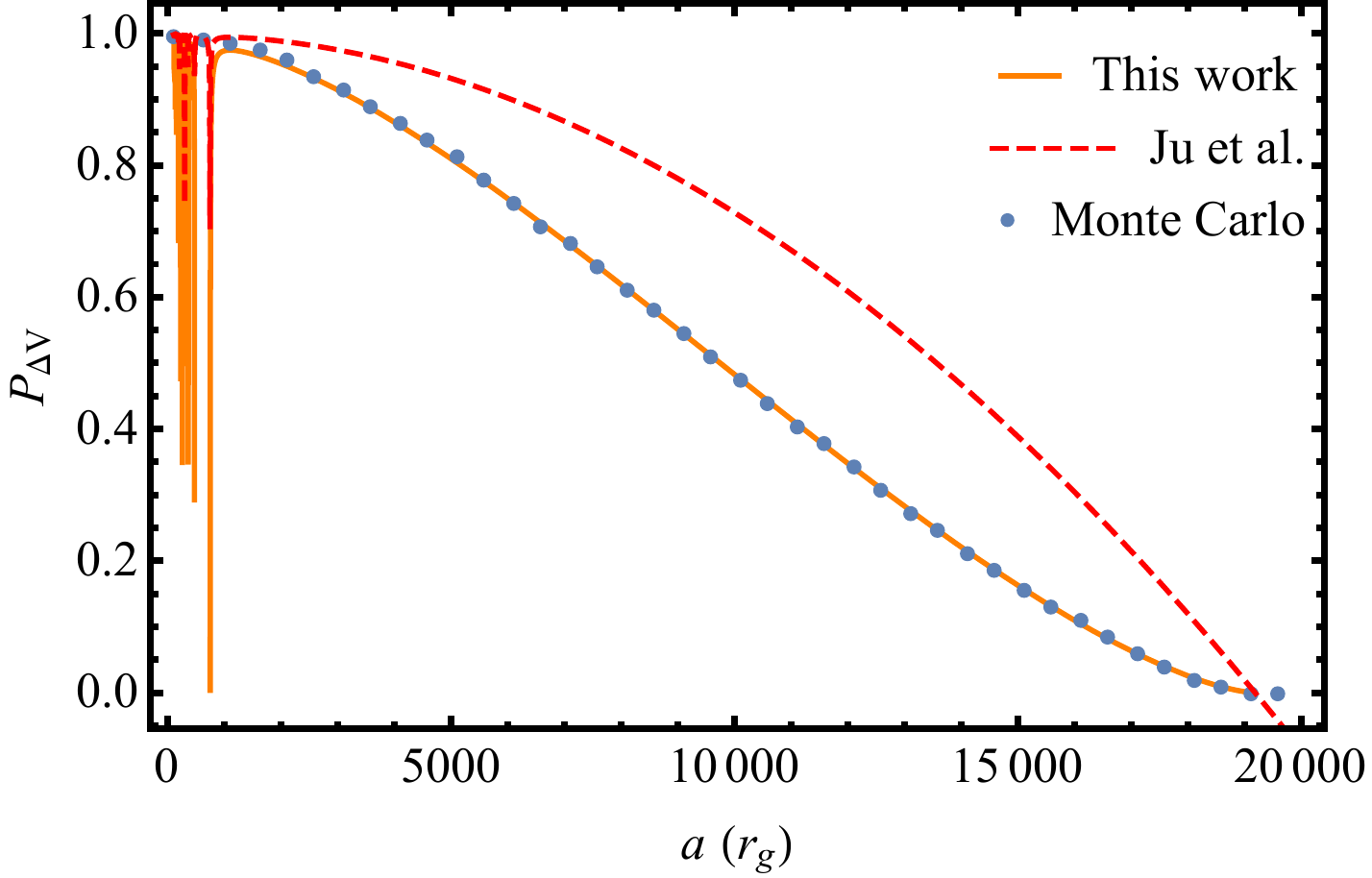} 
\end{center}
\caption{Comparison of the analytic solution for $P_{\Delta V}$ from equation~\ref{eq_PdV} (orange, solid line), Monte Carlo realization for the same range of SBHB orbital configurations (blue dots), and the solution presented in \citet[][red, dashed line]{Ju2013}. The probability is calculated for $M=10^8M_\odot$, $q=0.3$, $\Delta t = 1\,$yr, and the value of $\Delta V_{\rm lim}$ adopted in this model.}
\label{fig_compare}
\end{figure}

In order to verify the analytic solution for the probability distribution $P_{\Delta V}$ given in equation~\ref{eq_PdV}, we compare it with a Monte Carlo realization for the same range of SBHB scenarios and show the result in Figure~\ref{fig_compare}. The Monte Carlo calculation samples a large number of SBHB configurations, assuming a uniform distribution of $a$ and $q$, and about $10^4$ orbital orientations uniformly distributed in $\phi$ and $\cos \theta$. The two distributions are in agreement everywhere, except at small $a$, where the finite Monte Carlo sampling in $a$ does not capture the steep drops of the probability distribution to zero. Nonetheless, this result gives us confidence that our analytic solution correctly represents $P_{\Delta V}$ for a wide range of SBHB orbital separations detectable by spectroscopic surveys. We also compare our solution for $P_{\Delta V}$ to that presented in \citet{Ju2013} for completeness, and find that the two solutions differ. In our calculation we obtain the exact analytic solution for $P_{\Delta V}$ by setting up an integral with limits given by $\theta(\phi)$ and $\phi$, defined in equation~\ref{eq_angles2}. \citet{Ju2013} pursue somewhat different, approximate numerical approach to the integration which allows them to find an analytic expression for $P_{\Delta V}$ only if they make a simplifying assumption that $\theta$ and $\phi$ are independent functions. As illustrated  in Figure~\ref{overlap} however, $\theta$ and $\phi$ for which the criterion $\Delta V_{\rm obs} \geq \Delta V_{\rm lim}$ is satisfied are related and change in tandem. These differences account for the discrepancy between the two solutions illustrated in Figure~\ref{fig_compare}.

\section{D: Comparison of numerical and Monte Carlo solutions for $P_{V, \Delta V}$}
\label{app:Pvdv}

In \S~\ref{sec_Pvdv} we calculate the probability of detection by a distant observer of an SBHB that simultaneously satisfies the criteria $V_{\rm obs} \geq V_{\rm lim}$ and $\Delta V_{\rm obs} \geq \Delta V_{\rm lim}$ as
\begin{equation}
P_{V, \Delta V}=\frac{2}{\pi}\int_{\min(C,D)}^{\min(D,\pi/2)} d\phi \int_{0}^{\min(A,B)}\sin\theta \; d\theta \; ,
\end{equation}
where the minimum function is defined as $\min(A,B)=(A+B-|A-B|)/2$. Integrating over $\theta$ gives
\begin{equation}
P_{V, \Delta V}=\frac{2}{\pi}\int_{\min(C,D)}^{\min(D,\pi/2)}\left[1-\cos\left(\frac{A}{2}+\frac{B}{2}-\frac{|A-B|}{2}  \right)  \right] \; d\phi \; .
\end{equation}
Making use of the angle sum and difference trigonometric identities we get
%
%
\begin{equation}
P_{V, \Delta V}=\frac{2}{\pi}\int_{\min(C,D)}^{\min(D,\pi/2)}\left[1-\cos(\frac{A}{2}+\frac{B}{2})\cos\left({\frac{-|A-B|}{2}}\right)+\sin\left(\frac{A}{2}+\frac{B}{2}\right)\sin\left({\frac{-|A-B|}{2}}\right)  \right] \; d\phi
\end{equation}
%
%
 We can make further simplifications since we are dealing with only one octant, namely
%
\begin{equation}
\cos\left(-\left|\frac{A-B}{2}\right|\right)=\cos\left(\left|\frac{A-B}{2}\right|\right)=\cos\left(\frac{A-B}{2}\right)
\quad {\rm and}
\label{cosabs}
\end{equation}
%
\begin{equation}
\sin\left(- \left|\frac{A-B}{2}  \right|  \right)=-\sin\left( \left|\frac{A-B}{2}  \right|  \right)=-\left| \sin\left( \frac{A-B}{2}  \right)  \right|=-\sqrt{\sin^2\left( \frac{A-B}{2}  \right)} \; .
\label{sinabs}
\end{equation}

Using these expressions we rewrite the probability as
\begin{equation}
\begin{split}
P_{V, \Delta V}=\frac{2}{\pi}\int_{\min(C,D)}^{\min(D,\pi/2)}\Bigg[1-\cos^2\left(\frac{A}{2}\right)\cos^2\left(\frac{B}{2}\right) +\sin^2\left(\frac{A}{2}\right)\sin^2\left(\frac{B}{2}\right)\\
-\sqrt{\left(\sin^2\left(\frac{A}{2}\right)\cos^2\left(\frac{B}{2}\right)-\sin^2\left(\frac{B}{2}\right)\cos^2\left(\frac{A}{2}\right)   \right)^2}\Bigg] \; d\phi \; .
\end{split}
\end{equation}

We can further simplify the overall calculation by making use of the sine and cosine half-angle formulas:
%
%
%
\begin{equation}
P_{V, \Delta V}=\frac{2}{\pi}\int_{\min(C,D)}^{\min(D,\pi/2)}\left[1-\frac{\cos A}{2}-\frac{\cos B}{2}-\left|\frac{\cos B}{2}-\frac{\cos A}{2}   \right| \right] \; d\phi \; .
\label{eq_Pvdv}
\end{equation}

%
\begin{figure}[t]
  \centerline{
    \includegraphics[scale=.54]{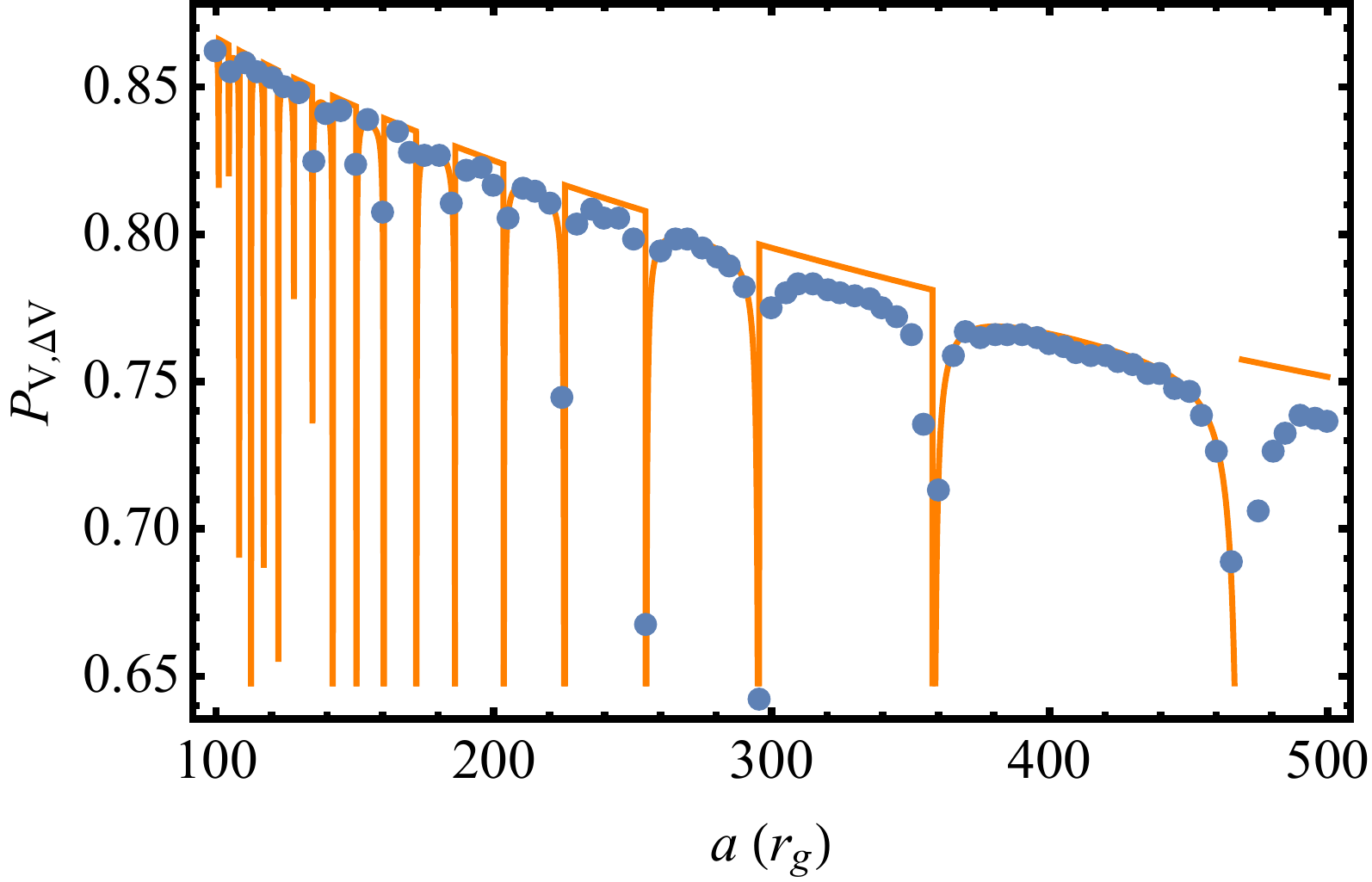}
    \hfill
    \includegraphics[scale=.52]{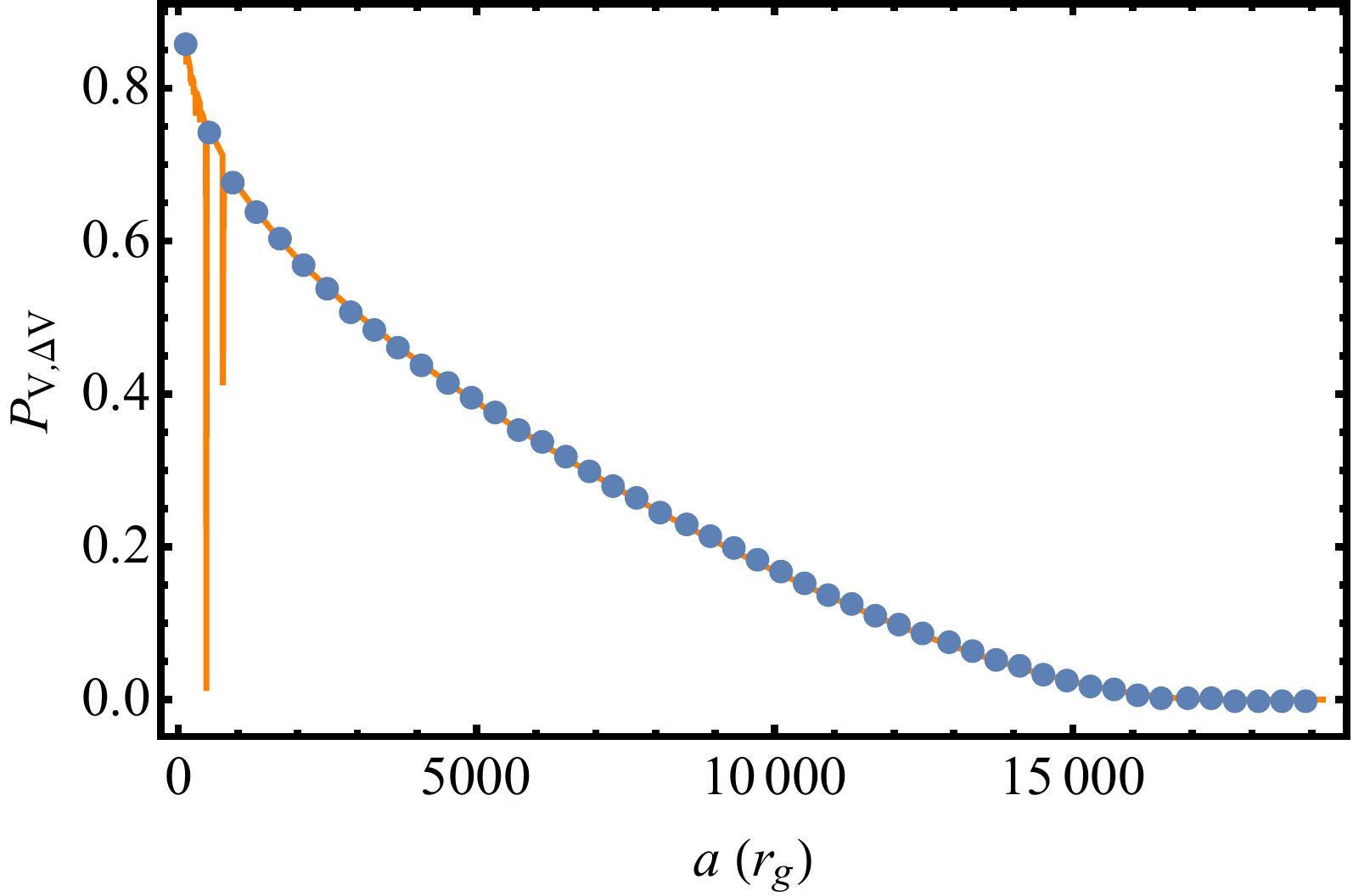}
    }
\caption{Comparison of the numerical solution for $P_{V, \Delta V}$ from equation~\ref{eq_Pvdv} (orange, solid line) and the Monte Carlo realization for the same range of SBHB orbital configurations (blue dots) showing a good agreement at the small (left) and large (right) orbital separations. The probability drops to zero at small $a$ where the cadence of observations corresponds to an integer number of SBHB orbital periods. The probability is calculated for $M=10^8M_\odot$, $q=0.3$, $\Delta t = 1\,$yr, and $V_{\rm lim}$ and the value of $\Delta V_{\rm lim}$ adopted in this model.}
\label{fig_Pvdv}
\end{figure}
We verify this solution by comparing it with a Monte Carlo realization of $10^4$ different orbital orientations determined by the angles $\theta$ and $\phi$ for a given orbital separation of an SBHB with $M=10^8M_\odot$ and $q=0.3$ (see Figure~\ref{fig_Pvdv}). Similarly to $P_{\Delta V}$, $P_{V, \Delta V}$ drops to zero at small $a$ where the cadence of observations happens to correspond to an integer number of the SBHB orbital periods. These instances are accurately traced by both the numerical and Monte Carlo solutions, albeit with only a few points in the Monte Carlo realization due to the finite sampling in $a$. Figure~\ref{fig_Pvdv} shows good agreement between the two solutions for both the small and large orbital separations, giving us confidence that equation~\ref{eq_Pvdv} correctly represents $P_{V, \Delta V}$.

\bibliographystyle{apj}
\bibliography{apj-jour,ref}


\end{document}